\documentclass[%
 aip,
 amsmath,amssymb,
 reprint,%
]{revtex4-1}

\usepackage{graphicx}% Include figure files
\usepackage{dcolumn}% Align table columns on decimal point
\usepackage{bm}% bold math
\usepackage[utf8]{inputenc}
\usepackage[T1]{fontenc}
\usepackage{mathptmx}
\usepackage{hyperref}

\usepackage{color}
\usepackage{subcaption}
\usepackage[textsize=tiny]{todonotes}

\usepackage{bmpsize}

\usepackage{empheq}

\usepackage{xspace}

\newcommand{\yso}[0]{Y$_2$SiO$_5$\xspace}

\newcommand{\tmyag}[0]{{Tm$^{3+}$:YAG}\xspace}
\newcommand{\perroothz}{$/\sqrt{\textrm{Hz}}$\xspace}

\begin{document}
\title{Piezospectroscopic measurement of high-frequency vibrations in a pulse-tube cryostat}
\date{\today }
\author{Anne Louchet-Chauvet}
\affiliation{Laboratoire Aim\'e Cotton, CNRS, Univ. Paris-Sud, ENS-Cachan, Universit\'e Paris-Saclay, 91405, Orsay, France}
\author{Rose Ahlefeldt}
\affiliation{Centre for Quantum Computation and Communication Technology, Research School of Physics and Engineering, The Australian National University, Canberra 0200, Australia}
\author{Thierry Chaneli\`ere}
\email{thierry.chaneliere@neel.cnrs.fr}
\affiliation{Laboratoire Aim\'e Cotton, CNRS, Univ. Paris-Sud, ENS-Cachan, Universit\'e Paris-Saclay, 91405, Orsay, France}
\affiliation{Univ. Grenoble Alpes, CNRS, Grenoble INP, Institut N\'eel, 38000 Grenoble, France}

\date{\today}
 
\begin{abstract}
Vibrations in cryocoolers are a recurrent concern to the end user. They appear in different parts of the acoustic spectrum depending on the refrigerator type, Gifford McMahon or pulse-tube, and with a variable coupling strength to the physical system under interest. Here, we use the piezospectroscopic effect in rare-earth doped crystals at low temperature as a high resolution, contact-less probe for the vibrations. With this optical spectroscopic technique, we obtain and analyze the vibration spectrum up to 700kHz of a 2kW pulse-tube cooler. We attempt an absolute calibration based on known experimental parameters to make our method partially quantitative and to provide a possible comparison with other well-established techniques.
% insert abstract here
\end{abstract}

% insert suggested PACS numbers in braces on next line

\maketitle

%\tableofcontents

\section{Introduction}

Compared to liquid helium cryostats, cryocoolers open many perspectives in the scientific community. First, closed-cycled systems allow to access cryogenic temperatures even when liquid helium is not available on-site. They can also be  operated continuously with very limited maintenance, which is mandatory in many integrated applications or when remote control is required~\cite{oh2013passive}. The price to pay is a higher level of vibration for the cryocoolers compared to their {\it wet} equivalents~\cite{radebaugh2009cryocoolers}. The measurement~\cite{tomaru2004vibration} and the reduction of the vibrations in a cryocooler are active subjects of research, finding applications in very different fields including metrology for gravitational wave detection~\cite{tomaru2005vibration}, the definition of high stability oscillators~\cite{grop2010elisa}, or as a routine instrument for fundamental research~\cite{caparrelli2006vibration, evans2008cryogen, wang2010vibration, stm, quacquarelli2015scanning, maisonobe2018vibration}.

These prospects have motivated many studies to accurately measure the vibration spectrum~\cite{caparrelli2006vibration, chijioke2010vibration} and to model it~\cite{riabzev2009vibration} up to few tens of kHz. The noise is essentially dominated by the low-frequency components driven by the compression cycle. Higher frequencies are also present because of high-order harmonics of the tubes' mechanical distortion modes and smaller parts of the apparatus with higher resonant frequencies. The high frequency region of the spectrum may have an impact on  micro or nano-mechanical resonators investigated at low temperatures to explore the quantum mechanical nature of massive objects~\cite{schwab2005putting} or cryogenic mirrors for gravitational wave observation~\cite{tomaru2005vibration}, since both could have resonances falling into the high frequency range.

More fundamentally, proper modeling requires to measure a large spectrum in order to design accordingly a passive stabilization scheme. Active stabilization offers an alternative and elegant solution~\cite{caparrelli2006vibration,majorana2006vibration}. This technique also requires the largest acquisition bandwidth to obtain a fast and reliable servo-control.

Accelerometers represent a commercial solution for measuring vibrations that are already compatible with cryogenic temperatures. They are routinely used to produce technical reports and allow a rapid comparison between different mechanical assembly of cryocoolers. When higher measurement bandwidths are targeted, optical techniques are particularly interesting. Fast response is guaranteed by current opto-electronics components, but they also offer a fine spatial resolution given by the beam size and an intrinsic nondestructive, contact-less character. Optical measurements are of two types: interferometric, when the surface under study reflects light as the end mirror of a Michelson interferometer, for example~\cite{mauritsen2009low, chijioke2010vibration}, or intensity sensor, usually based on fiber-coupled excitation and detection ports~\cite{caparrelli2006vibration}. Both give access to the kHz frequency range, up to 20kHz~\cite{chijioke2010vibration}, at least an order of magnitude better than most  accelerometers. The detection bandwidth of optical techniques could also be pushed to higher frequencies, away from the audible spectrum, which have been mostly unexplored despite the absence of clear technical limitations.%\todo{how could this be done/ why hasn't it been done?}
%{\it Why are there no optical measurements above 20 kHz ?}

We propose a new non-interferometric optical approach based on the piezospectroscopic effect in rare-earth doped oxide crystals at low temperature.  The technique has the same advantages as other optical measurements: high bandwidth and no electrical contact on the sample, this latter representing a possible source of thermal and mechanical perturbations.  The laser beam probes the stress induced by vibrations in the crystal instead of the change in distance between two reference points (as in, for example, interferometers). Piezospectroscopy has been used as a diagnostic tool for sapphire-based materials in which the inclusion of chromium can be used directly as a photoluminescent stress probe~\cite{he1995determination}. This {\it in-situ} method is particularly well-adapted to extreme fabrication conditions, as in high temperature reactors~\cite{christensen1996nondestructive, schlichting2000application}. It was also proposed in diamond NV-centers as a tool to detect the damage trail left by a weakly interacting massive particles (WIMPs)~\cite{rajendran2017wimp}. Piezospectroscopy is even more relevant at low temperatures because atomic optical transitions become narrower, giving an enhanced sensitivity. This transition narrowing can be enhanced in rare-earth doped solids with the spectral-hole burning (SHB) mechanism, whose spectral resolution is in the MHz range \footnote{as an order of magnitude, $400$~GHz corresponds to $1$~nm in the near IR}, well below the already narrow inhomogeneous linewidth at cryogenic temperature~\cite[chap.4]{liu2006spectroscopic}.

Here, we use the piezospectroscopic effect in rare-earth doped oxides as a diagnostic tool, but it was initially identified as a strong limitation to high resolution measurements based on SHB performed using cryocoolers. In recent years,  interest has arisen in using SHB for optical signal processing applications, such as  RF spectral analysis~\cite{merkel2004multi, gorju200510}, acousto-optic filtering~\cite{li2008pulsed}, quantum memories~\cite{heshami2016quantum}, quantum opto-mechanics \cite{Seidelin} and the definition of optical frequency standards~\cite{thorpe2011frequency}. Thus, there is a strong need for cryogenic coolers with low vibrations for these different applications. Custom solutions have been used to reduce the vibrations to a level compatible with the targeted spectral resolution~\cite{PhysRevB.94.075117, thorpe2013shifts, gobron2017dispersive}.

Beyond presenting a new technique for measuring vibrations in cryocoolers, our goal is to give a comprehensive analysis of the coupling between the vibrations and the rare-earth optical transition for the community studying these materials for signal processing applications. Our method could also establish a common testbed for different cryocoolers in order to compare the designs, upgraded or not, with custom isolation stages.

We first review the physics of piezospectroscopic measurements in rare-earth doped crystals. These measurements were an active subject of research in the early days of laser spectroscopy, when they were used to identify the local site symmetry of dopant impurity in solids under static pressure. We extend the static case to SHB materials, more specifically \tmyag (yttrium aluminium garnet), under dynamical pressure fluctuations (vibrations). We then introduce and use our piezospectroscopic method to characterize the vibration of a pulse-tube cryocooler. We obtain and analyze a vibration spectrum up to 700kHz and partially correlate our measurement with the direct acoustic recording of the rotary valve.

\section{Piezospectroscopy of rare-earth doped crystals}
The piezospectroscopic effect of rare-earth luminescent transitions was investigated in the 60's following the seminal work of Kaplyanskii~\cite{kaplyanskii1964noncubic, kaplianskii1965deformation}. Under applied pressure, rare earth levels shift and split, and this information can be used to determine the local site symmetry and thus understand how the  dopant rare-earth ion is included within the crystalline cell. The intrinsically narrow line of optical  4f-4f transitions makes the measurement of shifts and splittings possible even at room temperature, if a diamond anvil cell is used to apply tens of GPa.

We first briefly review static pressure measurements that involved different lanthanide dopants in oxide and fluoride crystals, commonly fabricated and used for lasers. We perform the same static measurement by employing the highly sensitive SHB technique in \tmyag giving access to much higher resolution and much lower applied pressure. We finally discuss the correspondence between the dynamical pressure fluctuations, namely the vibrations, and the static measurements.

\subsection{Static pressure measurements}

The application of a static pressure allows one to extract information about the local crystal field surrounding the luminescent center. We briefly and partially review a few results of piezospectroscopic measurements in rare-earth doped oxydes and fluorides (see Table~\ref{table:review}).

\begin{table}[h]
\begin{center}
\begin{tabular}{ |c|c|c|c|c| }
 \hline
Refs & Ion$^{3+}$ & Transition & Crystal & Shift (Hz/Pa) \\ \hline\hline
\cite[Table III]{bungenstock2000effect} &Pr& $^{3} H_{4} \rightarrow $ avg. & LaOCl &-242.92 \\ \hline
\cite[Table I]{manjon2001effect} & Nd & $^{4}F_{3/2}(1) \rightarrow $$^{4}I_{9/2}(1)$ &YLiF$_4$ & -37 \\ \hline
\cite[Table 1]{troster2003crystal} & Pr & $^{3}H_{4}(1) \rightarrow $$^{1}D_{2}(1)$ &YLiF$_4$&  -172 \\ \hline
\cite[Table I]{manjon2004effect} & Nd & $^{4}F_{3/2}(1) \rightarrow $$^{4}I_{9/2}(1)$ &YVO$_4$ & -193 \\ \hline
\cite[fig.1]{rodriguez2006high}& Nd & $^{4}F_{3/2} \rightarrow $$^{4}I_{9/2}$ &LiNbO$_3$ & $\sim$ - 270  \\ \hline
\cite[fig.1a]{turos2007pressure} & Pr & $^{3}H_{4} \rightarrow $$^{1}D_{2}$ & YAG & $\sim$ -100\\ \hline
\cite[fig.3]{thorpe2011frequency} & Eu & $^{7}F_{0} \rightarrow $$^{5}D_{0}$ (site 1) & \yso & -211\\ \hline
\cite[fig.3]{thorpe2011frequency} & Eu & $^{7}F_{0} \rightarrow $$^{5}D_{0}$ (site 2) & \yso & -52\\ \hline
\cite[fig.7a]{kaminska2016spectroscopic}&Yb & $^{2}F_{7/2}(1) \rightarrow $$^{2}F_{5/2}$(1) & GGG& 76 \\ \hline
\end{tabular}
\end{center}
\caption{Review of piezospectroscopic measurements in rare-earth doped crystals. GGG stands for gadolinium gallium garnet. $^{4}F_{3/2}(1)$ means, for example, the first (lowest) crystal field level of the $^{4}F_{3/2}$ multiplet. Site 1 or 2 are two crystallographic sites of \yso \cite{thorpe2011frequency}.}
\label{table:review}
\end{table}

A wide range of dopants, transitions and host matrices has been studied in the references~\cite{bungenstock2000effect, manjon2001effect, troster2003crystal, manjon2004effect, rodriguez2006high, turos2007pressure, kaminska2016spectroscopic}, and in Table~\ref{table:review} we give some representative values.
Despite very different combinations of transitions and crystals, the piezospectroscopic coefficients (in Hz/Pa) have very similar values and are dispersed by less than an order of magnitude. This similarity is related to the structural resemblance of the crystals. The piezospectroscopic effect is an indirect Stark shift induced by a modification of the crystalline electric field surrounding the rare-earth dopant. The applied pressure translates into a compression (strain) which brings the ionic charges closer. The dopant transitions are then shifted due to their sensitivity to electric fields (the Stark effect). The linear Stark shift in rare earth 4f-4f transitions varies only by about two orders of magnitude across different ions and host materials \cite[Table 2]{macfarlane2007optical}. The variation in Young's modulus and inter-atomic distances is even smaller, resulting in comparable Stark coefficients across very different materials.

Meanwhile, lanthanide inter-shell 4f-5d transitions have much stronger piezospectroscopic coefficients, as for example 4200 Hz/Pa for Pr$^{3+}$ in YAG~\cite{meltzer2005pressure}. The reason is that the 4f orbitals of the lanthanides are  inner shells benefiting from a lower sensitivity to environmental changes, including the piezospectroscopic effect. In terms of measurement technique, the lower sensitivity of the intra-shell 4f-4f transitions is nevertheless largely compensated by the much narrower linewidth when probed by narrowband lasers. The resolution is, in that case, not limited by the inhomogeneous broadening (few GHz), but the homogeneous linewidth (below MHz) when the SHB technique is used, as we will see in section~\ref{shb_tmyag}.

In summary, a typical value of 100 Hz/Pa can be kept in mind to evaluate the order of magnitude of pressure-induced shifts. As we will see now, the $^{3} H_{6} \rightarrow $$^{3}H_{4} $ transition of \tmyag at our wavelength of interest (793nm) falls in this range as well.

\subsection{Piezospectroscopy of \tmyag}\label{shb_tmyag}
The SHB technique has been extensively used to measure the homogeneous linewidth of many rare-earth transitions and has been described in many textbooks, see~\cite[chap.4]{liu2006spectroscopic} for example. Put simply, the technique involves using a narrow band laser to select out a spectrally narrow portion of a broader atomic line by pumping atoms into a non-resonant shelving state. The width of this artificially narrow line (the spectral hole) is determined by the linewidth of the laser, well below MHz for narrowband lasers (such as extended cavity diode lasers). For very narrow lasers, the resolution is eventually limited to a few kHz,   the homogeneous broadening of rare-earth transitions such as those in \tmyag. The lifetime of the spectral hole can be extremely long, hours or days, by applying a field and shelving population in the  Zeeman hyperfine structure~\cite{ohlsson2003long, deseze}. The SHB technique has been refined to the extreme in the context of optical metrology~\cite{thorpe2011frequency, gobron2017dispersive} and used to accurately measure the detrimental shifts induced by the environmental changes, including pressure shifts \cite{thorpe2013shifts}.

\begin{figure}[h]
    \centering
    \includegraphics[width=0.7\columnwidth]{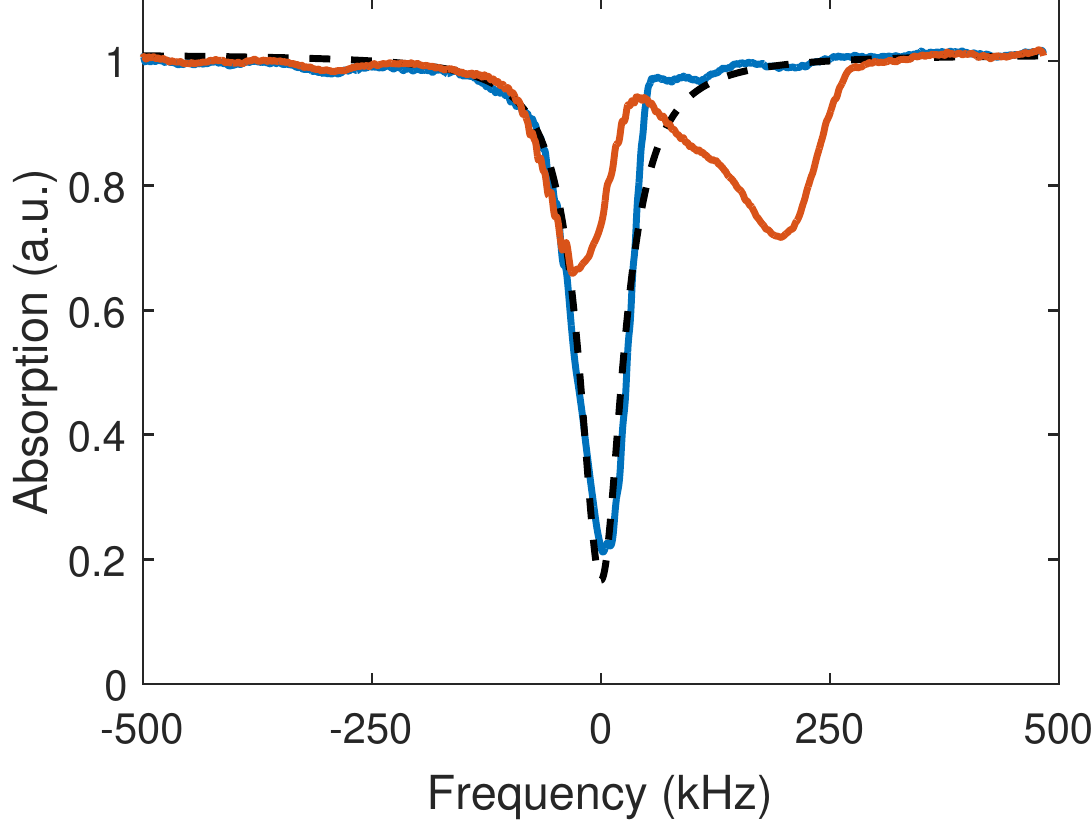}
    \caption{Example of SHB absorption spectrum in \tmyag (blue line). The hole has a roughly Lorentzian shape (black dashed line) of 50kHz width (FWHM). When a 3.3kPa pressure is applied along the [110] crystal axis, the line splits in two components (red line) that shift (and broaden) independently (see text for details).}
    \label{fig:shb_ex}
\end{figure}

In Fig.\ref{fig:shb_ex}, we show the effect of a small applied pressure on the shape and position of a spectral hole on the $^{3} H_{6} \rightarrow $$^{3}H_{4} $ transition of \tmyag. The initial hole (without pressure) is $50$~kHz broad, significantly larger than the homogeneous linewidth ($\sim$3kHz) because of power broadening. This resolution is sufficient to observe the dramatic effect of even a few kPa.

A splitting of the line is actually expected when a pressure is applied along the [110] crystal axis in YAG \cite{kaplyanskii1964noncubic, kaplianskii1965deformation}. Tm substitutes for Y in the YAG cell at a site of D$_2$ symmetry. This site has 6 possible orientations relative to the crystallographic axes, numbered from 1 to 6 (Ref. \onlinecite[Fig. 1]{sun2000symmetry} and references therein). Five of these sites contribute to the spectral hole in Fig. \ref{fig:shb_ex}, site 1, and sites 3 to 6. Site 2 does not, because the optical transition is polarized, and for site 2 this polarization is parallel to the beam propagation direction, so the site is not excited. For pressure along [110], sites 3 to 6 are equivalent, and shift as a group in one direction, while site 1 shifts in the other, resulting in the split line in Fig. \ref{fig:shb_ex}. Which peak corresponds to which site was determined by looking at the polarisation dependence of the split peaks.

The goal of our paper is not to describe extensively the piezospectroscopy effect in \tmyag, but rather to give a general idea of the link between the pressure vibrations and the optical transition. The measurement
in Fig.~\ref{fig:shb_ex} is illustrative and gives an order of magnitude of the pressure shifts. It was obtained in a dedicated variable temperature insert (VTI) cryostat (2-3K) that was used to apply a calibrated pressure and perform the piezospectroscopic analysis of our \tmyag sample. The shift in Fig.~\ref{fig:shb_ex} was measured as a function of the applied pressure. To vary the pressure, we used a piezoelectric actuator (Thorlabs AE0203D04F) that was inserted in the cryostat and constrained in contact with the crystal. To calibrate the piezo response, which is unknown at low temperature, we applied a static pressure by loading, with a calibrated weight, an inner rod that presses on the crystal though the VTI sample holder. A few volts across the piezo corresponds to $10$~kPa (a few tens of grams on the crystal surface).

\begin{figure}[h]
    \centering
    \includegraphics[width=0.99\columnwidth]{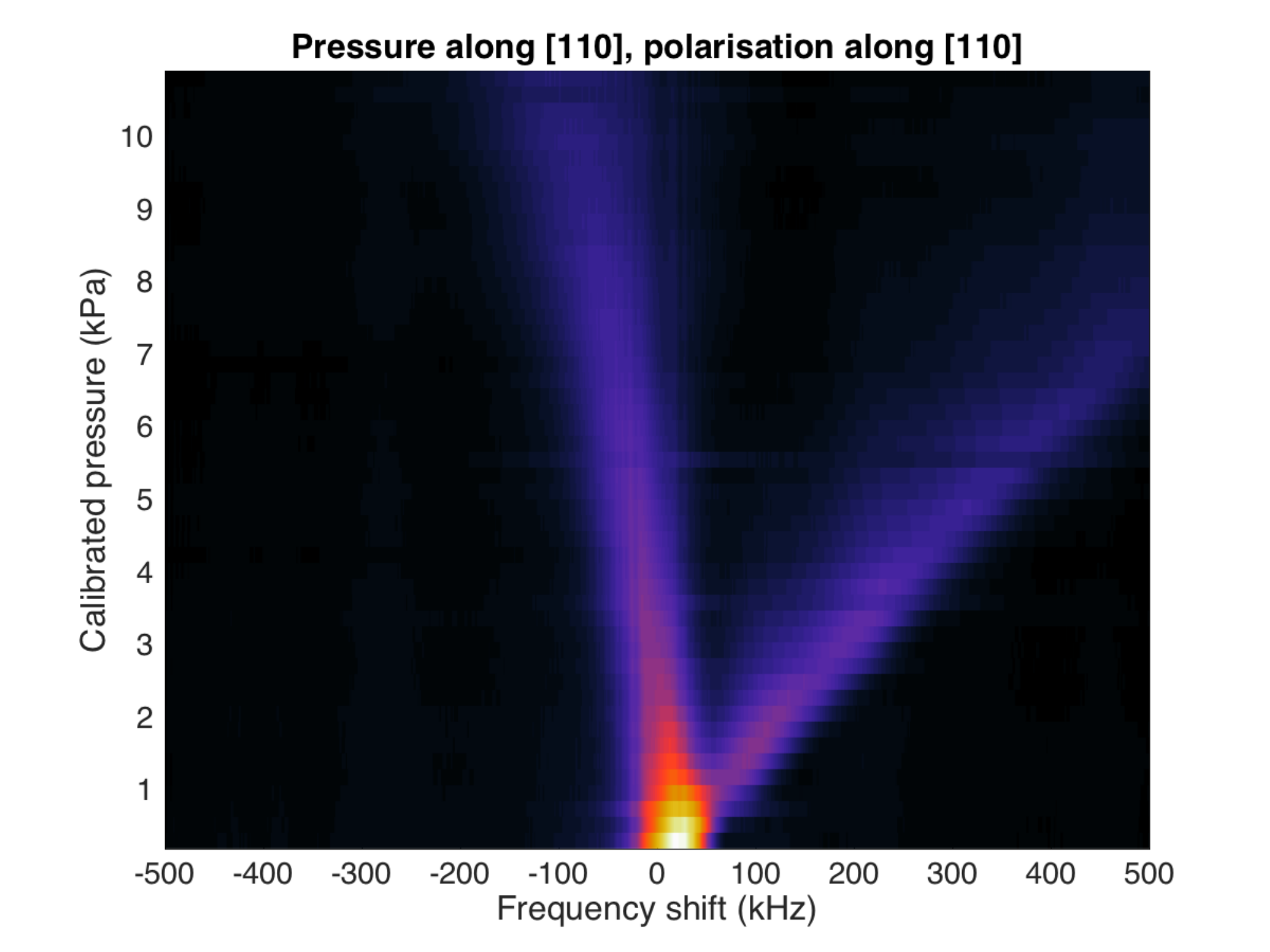}
    \caption{Pressure shifts of the different crystallographic sites when the pressure is applied along [110] The laser polarisation is along [110] as well, and so addresses only sites 1 and 3 to 6. Site 1 causes the peak with large positive shift $68$~Hz/Pa, and sites 3 to 6 the peak with a small negative shift $-13$~Hz/Pa.}
    \label{fig:shb_tmyag}
\end{figure}

We can see  in Fig.~\ref{fig:shb_tmyag} that the lines splits with different pressure coefficients for the different crystallographic sites. The splittings as a function of the piezo calibrated voltage are linear and allow us to extract a $68$~Hz/Pa coefficient for site 1 and $-13$~Hz/Pa to sites 3, 4, 5 and 6.  More generally, we observed shifts ranging from 13~Hz/Pa to 150~Hz/Pa depending on the sites and on the direction of applied pressure, demonstrating the anisotropic character of the piezospectroscopic response in Tm:YAG. Nevertheless, a complete analysis of the piezospectroscopic tensor for the different sites is beyond the scope of this paper. In the following, we will keep the value of 68~Hz/Pa as a reference because it is somehow intermediate amongst the various measurements made. We first remark that \tmyag has a similar shift to other rare earth materials (see Table~\ref{table:review}). We note that the line showed substantial broadening as well as splitting, indicating that not all ions in the crystal experienced the same pressure. This shows that it is difficult, even in a well-controlled measurement, to apply a uniaxial pressure that only shifts and does not broaden the line. We attribute the large broadening (with respect to the shift) to the inhomogeneity in the local strain field. Because of the surface roughness, the force may concentrate on few contact points. The spread from the surface to the interior of the crystal requires a specific mechanical analysis. We indeed discard any inhomogeneity of the local microscopic piezospectroscopic tensor (local distortion of the crystal cell) which could also lead to a broadening even under a uniform strain field. The local distortion of the crystal cell can be estimated by comparing the optical inhomogeneous line broadening (10~GHz$\sim$0.3~cm$^{-1}$) and the typical crystal field splitting (a few 100~cm$^{-1}$). So we expect the microscopic piezospectroscopic tensor to be homogeneous at less than the percent level. This estimation indicates by default that the broadening is due to the inhomogeneity of internal strain field and not the piezospectroscopic tensor.
In addition, the presence of multiple site orientations means that an applied pressure would shift all the sites differently.  These two features combined (intrinsic and multi-site broadenings) mean that the vibrations in a cryocooler are most likely to broaden the spectral line, with the amount of broadening as a measure of the strain in the crystal. Thus, we will use the measured piezospectroscopic coefficient $\kappa=68$~Hz/Pa as representative of the line broadening in \tmyag.

\subsection{Piezospectroscopy and vibrations}\label{piezo_and_vib}
\begin{figure}
    \centering
        \includegraphics[width=0.99\columnwidth]{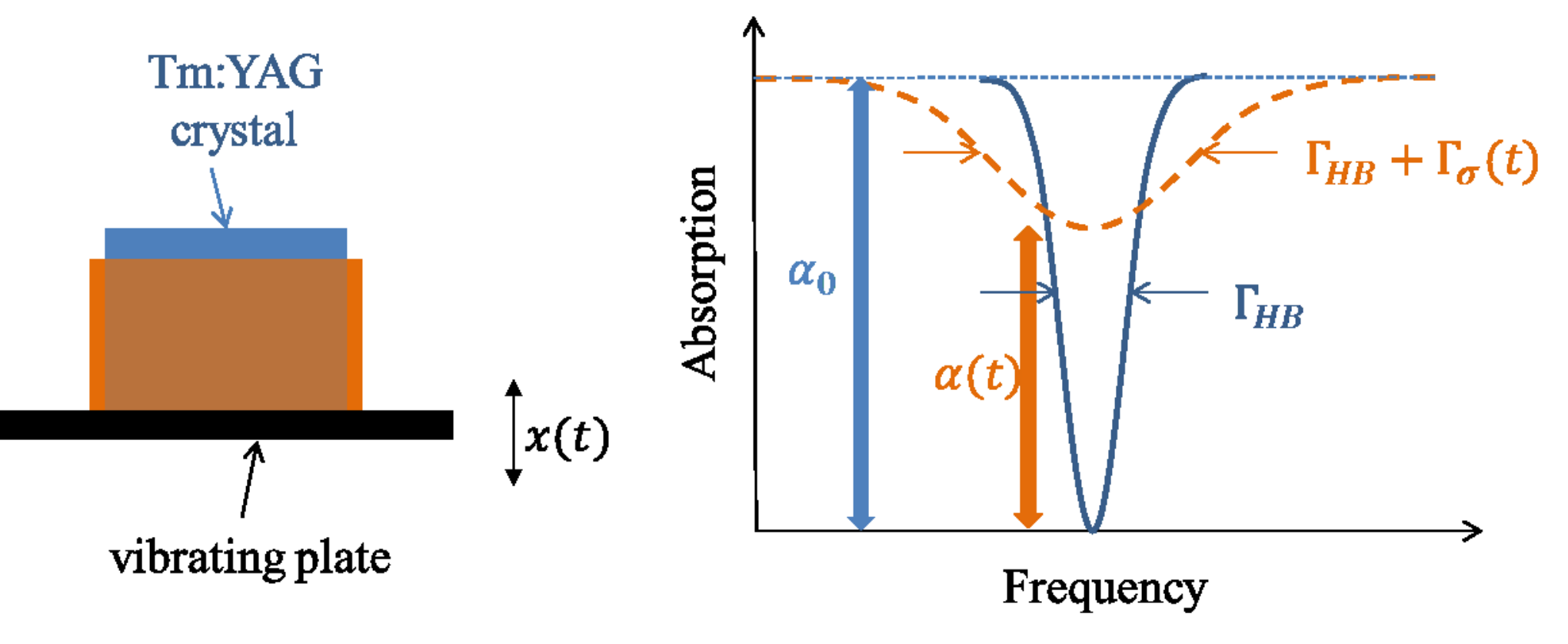}
    \caption{Toy model relating the vibrations and the piezospectroscopic measurement. Left: the vibrating plate (along $x$) induces a compression of the crystal (unconstrained in blue and compressed crystal in orange). Right: We assume that the compression broadens the spectral hole (from the solid blue to the dashed orange curve) and therefore reduces the hole size and increases the measured absorption in the center of the hole due to hole area conservation.}
    \label{fig:toy_model}
\end{figure}

Above we described the physical origin of the piezospectroscopic effect and illustrated the discussion with a static pressure measurement in \tmyag. The extension of this work to the case of vibrations, as dynamical pressure fluctuations, is not direct and requires some modeling. We propose a toy model in which the sample is attached to a vertically vibrating plate (see fig.~\ref{fig:toy_model})

The piezospectroscopic effect is due to the compression of the crystal when pressure is applied. This latter is related to the strain by $\sigma= E \displaystyle \Delta L /L$ where $E$ is the Young's modulus and $\frac{\Delta L }{L}$ the relative compression ($L$ is the sample length along $x$). In our model, the crystal is attached at one end where the vibration are induced as $x\left(t\right)$.
For a unidirectional propagation of the vibration, the compression is due to the retarded propagation of sound in the material, so quantitatively, $\Delta L = x\left(t\right) - x\left(t-L/V\right)$ where $V$ is the sound velocity in the solid.
For acoustic waves below $1$~MHz, the wavelength is much longer than the crystal so we can write $\displaystyle \frac{\Delta L }{L}=\frac{\dot{x}\left(t\right)}{V}$ or for the strain $\displaystyle  \sigma=\frac{E}{V}\dot{x}\left(t\right)$ to the first order.

The relation between the strain induced in the sample and the atomic line shape is not trivial. As we discussed in \ref{shb_tmyag}, a uniaxial pressure should in principle split the different sites.
As we have seen, obtaining a pure shift without broadening is quite challenging in the static case because of the effective inhomogeneity of the strain field in the crystal along the axis of the laser. In the dynamical case, we expect the strain field to be even less homogeneous so the line is essentially broadened by the vibration, summing the effect of ions in different locations in the crystal and of each crystallographic site. The line broadening $\Gamma_\sigma=\kappa |\sigma|$  can then be related to the crystal velocity:
\begin{equation}
\Gamma_\sigma^\mathrm{[Hz]} \left(t\right) =\kappa \frac{E}{V}|\dot{x}\left(t\right)| = 2.5\times10^9 \, |\dot{x}^\mathrm{[m/s]}\left(t\right)|
\label{eq:gamma}
\end{equation}
where the numerical example corresponds to YAG with $E=300$~GPa and $V=8165$~m/s. The value of $\kappa=68$~Hz/Pa is inferred from the spectral hole shift measurements reported in section~\ref{shb_tmyag}.

The absolute value of the velocity in Eq.~\ref{eq:gamma} makes it mathematically impossible to determine the signed velocity. An algorithmic retrieval of the signed signal is still possible but is a complex signal processing task \cite{discrete_sign} that we do not consider in a first approach. To proceed further with the calculation we are then compelled to crudely removing the absolute value, which comes down to assuming that negative strain leads to a narrowing of the spectral hole or that there is no negative strain at all. We expect this simplification to distort the sample displacement spectral density, with a possible doubling of the dominating frequencies, and the appearance of high frequency artefacts in the case of sharp features close to zero. The displacement spectral density will therefore be analyzed with caution.

The linewidth broadening induced by the vibration can be estimated optically by probing the absorption at the center of the hole as sketched in Fig.~\ref{fig:toy_model} (right) and as will be discussed in more details in \ref{transmission}. %The reader who is not familiar with the SHB technique may be confused by the inverted Lorentzian in Fig.~\ref{fig:toy_model} \todo{In all the other figures, you have positive holes. you could just stick with this}. This is the essence of hole-burning leading to an optical bandpass filter instead of bandstop for an absorbing line.
%Following the Bouguer-Beer-Lambert's absorption law, the logarithm of the transmission is proportional to the absorption coefficient.
A line broadening (for example by a factor of $3$ in Fig.~\ref{fig:toy_model}, right) increases the absorption in the center of the spectral hole accordingly.

We can write the hole linewidth in terms of the measured absorption at the center of the spectral hole, and the original spectral hole width $\Gamma_\mathrm{HB}$ at the instant of the time-resolved measurement. $\Gamma_\mathrm{HB}$ is not obviously the hole width in absence of vibration but it can be the result of a complex hole burning dynamics. It will be measured experimentally.  The vibration makes the hole width increase up to $\Gamma_\mathrm{HB}+\Gamma_\sigma(t)$, leading to a varying absorption coefficient $0\leq \alpha(t)\leq \alpha_0$ where $\alpha_0$ is the absorption in the absence of the spectral hole. Assuming a spectral hole broadening as previously discussed, and the conservation of the hole area, we write:
\begin{equation}
[\alpha_0 -\alpha(t)][\Gamma_\mathrm{HB}+\Gamma_\sigma(t)] =\alpha_0\Gamma_\mathrm{HB}
\label{eq:alpha}
\end{equation}
leading to :
\begin{equation}
\Gamma_\sigma(t)=\frac{\alpha(t)}{\alpha_0 -\alpha(t)}\Gamma_\mathrm{HB}
\label{eq:alpha2}
\end{equation}
This simply reflects the conservation of the absorbing centers involved in the SHB process, which is valid as long as the hole is measured over times shorter than the hole decay time.

%We now explain how we derive the velocity $\dot{x}(t)$.

As a conclusion, by using a simple model, we can relate the vibration (Eq.\ref{eq:gamma}) to the time-resolved SHB spectroscopy of $\alpha\left(t\right)$ (Eq.\ref{eq:alpha}) and finally extract quantitatively the characteristic vibration through its velocity $\dot{x}\left(t\right)$.

\section{Measurement setup}
We used a pulse-tube cryocooler (TransMIT PTD-009) with an Oerlikon COOLPAK 2000A compressor. The rotary valve is rigidly  attached to the cold head and has a cycle rate of $2$~Hz. A sound level meter (RadioShack 33-099) was positioned in contact with the outer shield of the rotary valve and records the audio signal in the 1Hz--20kHz range.

A copper inner vacuum chamber is attached to the second stage at $2.9$~K. It will be used for further isolation with exchange gas injection in the future but can simply be considered as a rigid sample mount for this experiment. The \tmyag sample was  resting at the bottom of the chamber, thermally contacted by Apiezon-N grease. This should ensure a rigid mechanical contact at low temperature (below 236~K, the glass transition temperature Apiezon-N \cite{KREITMAN197232}). The thermal contact is clearly maintained during the cooling cycle (absence of laser heating) so we can safely consider that the sample is rigidly contacted through the grease layer on cold stage.

The crystal was fabricated by Scientific Materials and is anti-reflection-coated on both sides to reduce interferometric effects.
The laser propagated along the $[1\bar{1}0]$ crystallographic axis. The thulium concentration in the sample is $0.25$ at.$\%$ so that the total absorption is $\alpha L\simeq2$ for $L=5$~mm. Further details about SHB in \tmyag including the population dynamics under optical excitation can be found in \cite[and references therein]{Ahlefeldt_PhysRevB.92.094305}.

The narrow-band laser source (below 1kHz) was an extended cavity diode laser stabilized to a Fabry-Perot cavity via a Pound-Drever-Hall feedback loop and tuned close to the center of the \tmyag $^3H_6(1)\rightarrow^3H_4(1)$ absorption line ($12604.42$~cm$^{-1}$). Diode lasers offer a moderate level of power fluctuations that could be interpreted as frequency shifts in our case. The use of a well-isolated reference locking cavity has the advantage to avoid the coupling of the crycooler's vibrations and laser cavity through air or optical table which would induce an acoustic frequency noise of the laser. This would be incorrectly interpreted as a piezospectroscopic shifts.
The laser beam was spectrally and temporally shaped with an acousto-optic modulator (AA-Optoelectronics MT110) driven with an arbitrary waveform generator (Tektronix AWG5004).
The $60~\mu$W beam was focused into the sample with a $100~\mu$m waist, corresponding to a Rabi angular frequency around $100$~krad.s$^{-1}$. We detected the transmitted light through the crystal with an avalanche photodiode (Thorlabs APD110A). The photodiode and the soundmeter signals were acquired via a digital oscilloscope (Agilent DSO5034A) as a $2$~s long trace containing 4 million points per channel, leading to a $2$~MHz sampling frequency.

\begin{figure}
    \centering
    \includegraphics[width=0.99\columnwidth]{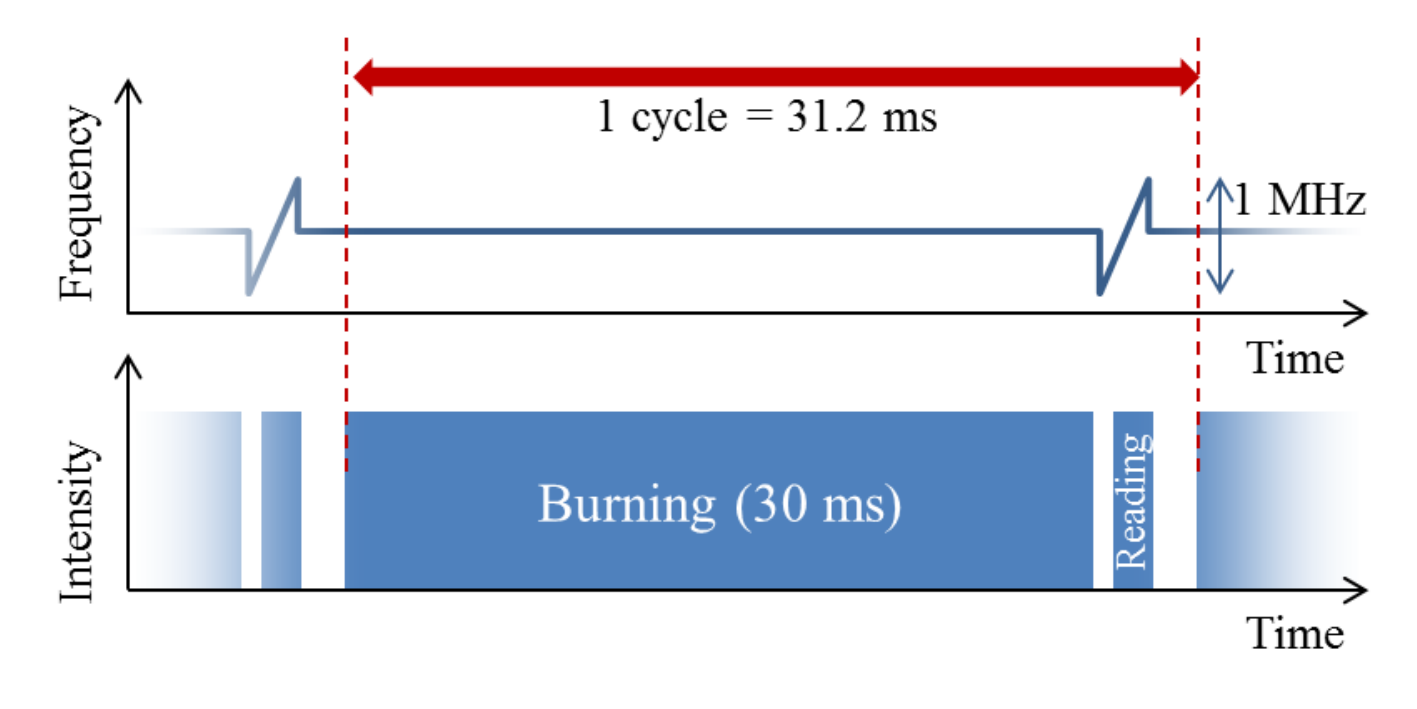}
    \caption{Excitation pulse sequence. The burning pulse is monochromatic and lasts 30~ms, whereas the reading pulse is frequency chirped over $1$~MHz in $200~\mu$s. Both pulses have the same power.}
    \label{fig:sequence}
\end{figure}

In an experiment, the hole amplitude at line center can be directly measured by monitoring the absorption of a laser burning a continuous spectral hole. A continuous measurement like this allows us to measure up to very high frequencies, limited only by the timing resolution of the acquisition system. However, it is still necessary to measure $\Gamma_\mathrm{HB}$, the spectral hole width, by chirping the laser over the hole to measure the hole shape. In the continuous scheme, this reading chirp must be done periodically, at an interval of $31.2$~ms in our case, comparable to the lifetime of the spectral hole, to provide an accurate reference.

The periodic pulse sequence depicted in Fig.~\ref{fig:sequence} was repeated with a $31.2$~ms cycle time, \emph{ie} around 16 times per rotary valve cycle.
The $30$~ms burning pulse filled $96\%$ of the cycle time and can therefore be regarded as almost continuous. The hole decay time was typically $10$~ms (limited by the lifetime of the Tm metastable state $^{3}F_{4}$) so each burning-reading step can be considered as independent.
The $100~\mu$s reading pulse was linearly chirped around the burning frequency so as to provide a measurement of the hole shape.%, 16 times per rotary valve cycle.

\begin{figure}
    \centering
    \includegraphics[width=\columnwidth]{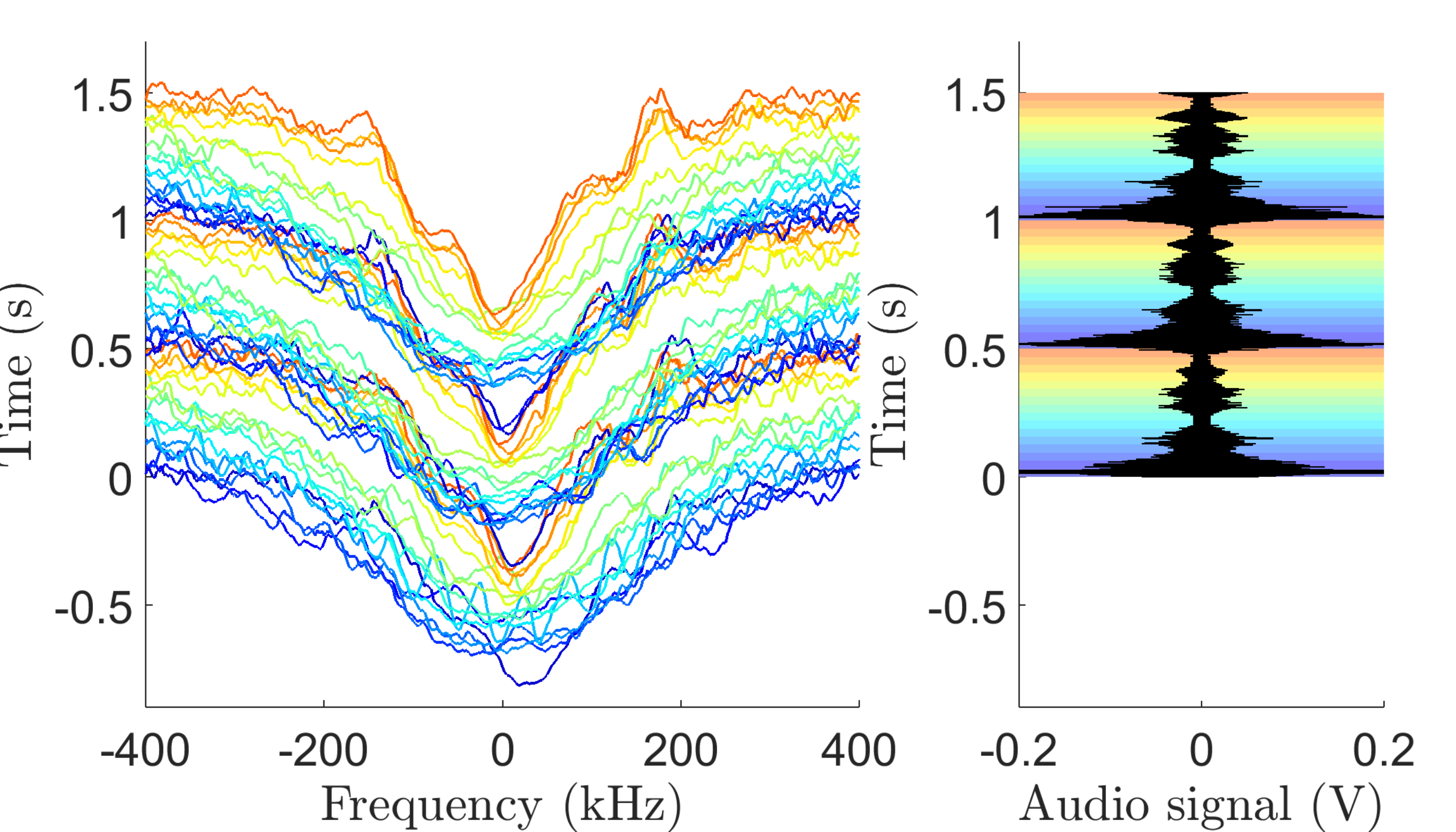}
    \caption{Left: Spectral hole absorption throughout 3 rotary valve cycles (\emph{ie} 48 sequence cycles). The graphs are vertically offset for a clear view of the spectral hole evolution with time. Right: the audio signal is simultaneously recorded (black line). Each sequence cycle is identified with a color (from blue at the beginning of the rotary valve cycle to orange at the end).}
    \label{fig:SpectresHB}
\end{figure}

We first plot the hole shape, measured by the reading chirped pulse, throughout 3 rotary valve cycles together with the audio level in Fig.~\ref{fig:SpectresHB}. The FWHM of the spectral hole oscillates between $150$~kHz and $350$~kHz depending on the position of the scan in the rotary valve cycle. This broadening is the result of the $30$~ms burning pulse under vibration. More specifically, the spectral hole is narrowest at the beginning of the cycle (that we define as the time where the sound level is maximum), then suddenly broadens after $60$~ms, remains broad for around $0.3$~s and then gets narrow again for the remaining $0.2$~s of the rotary valve cycle.

\section{Data analysis}

\subsection{Monitoring the vibration via the atomic absorption}\label{transmission}

The burning pulse both burns and probes the hole, providing a continuous readout of the hole amplitude. The read pulse periodically measures the entire hole shape, and is used as a reference for the hole width $\Gamma_\mathrm{HB}$ inferred from the amplitude measurement of the burn pulse.
More precisely, according to Bouguer-Beer-Lambert's absorption law, the transmission signal collected on the avalanche photodiode during the burning pulse is proportional to $e^{-\alpha(t)L_0}$ ($L_0$ is the laser propagation distance), eventually yielding the hole instantaneous broadening $\Gamma_\sigma(t)$ due to the vibrations, using Eq.~\ref{eq:alpha2}. The reference hole width $\Gamma_\mathrm{HB}$ is derived from the hole spectrum measured immediately after a given burning pulse in the repeated sequence of Fig.~\ref{fig:sequence}. We insist on the fact that $\Gamma_\mathrm{HB}$ is not the hole width in absence of vibration. It is the hole width at the moment when $\alpha(t)$ is monitored. %\todo{in Eq \ref{eq:alpha2} you talk about using the absorption at the center to measure the broadening, but you actually measure the whole lineshape anyway with the chirp. So why not just use the linewidth directly?}

\begin{figure}
    \centering
    \includegraphics[width=0.99\columnwidth]{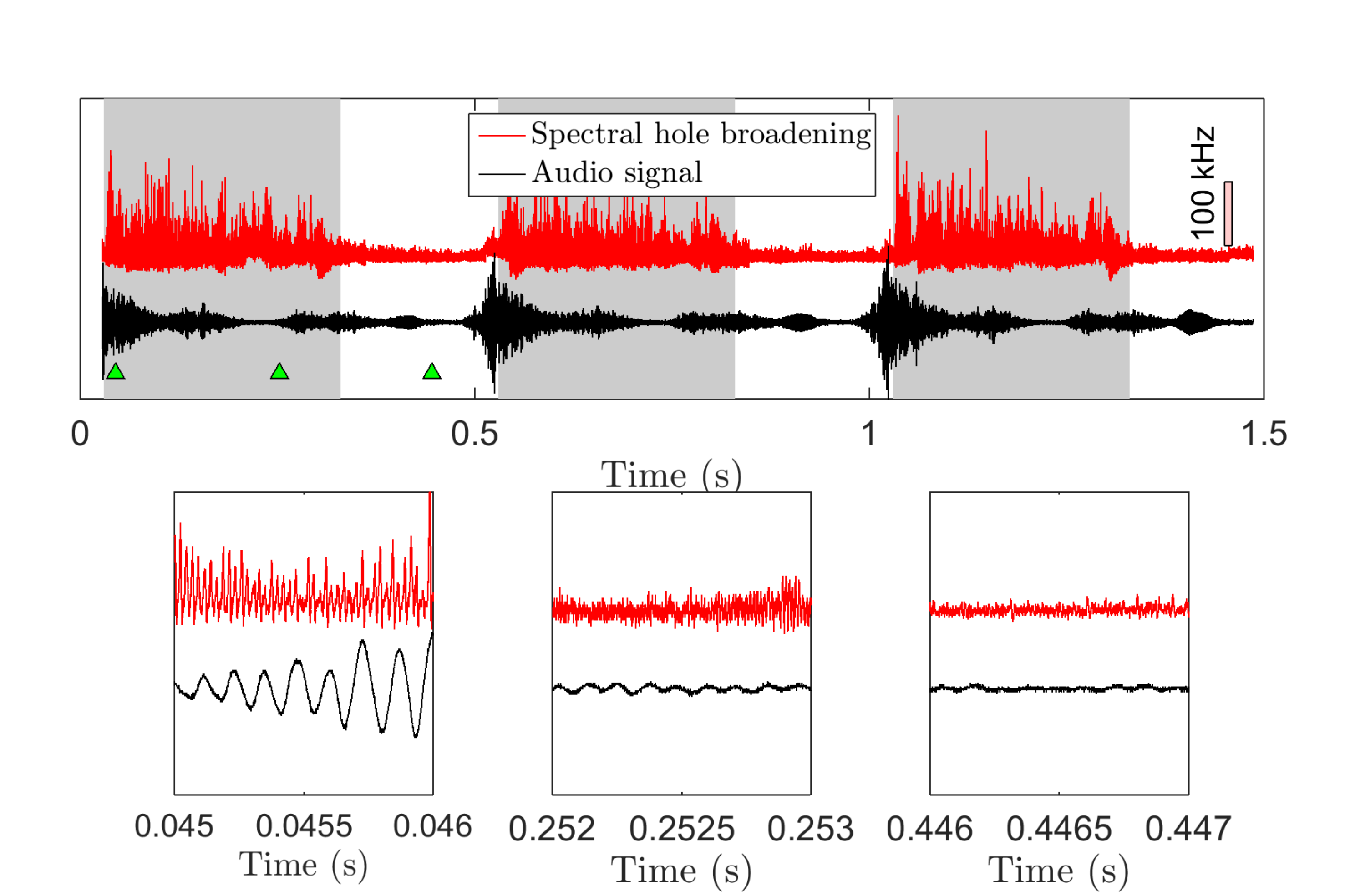}
    \caption{Experimental traces of the sound level $V_{ac}(t)$ (black) and the spectral hole broadening $\Gamma_\sigma(t)$ (red) deduced from the absorption measurement through the spectral hole during the burning sequence. The rectangle on the right gives the scale for the hole broadening trace. The $0.3$~s wide gray areas represent the so-called ``noisy'' phase defined in Sec.~\ref{sec:spectrotemporal}. The three lower graphs show the same traces on a 1 ms interval at times indicated by a green triangle on the main graph, corresponding to different positions in the rotary valve cycle. }
    \label{fig:TracesBrutes}
\end{figure}

A typical example of the hole broadening $\Gamma_\sigma(t)$ and of the sound level $V_{ac}(t)$ measured simultaneously are given in Fig.~\ref{fig:TracesBrutes}. We observe that the hole broadening oscillates at frequencies above 20~kHz and its value often exceeds $100$~kHz in the first $0.3$~s of each rotary valve cycle. The spectral hole perturbations are slightly delayed with respect to the rotary valve cycle. The acoustic noise and the piezospectroscopic perturbation have a common origin and are both triggered at the same time in the compression cycle. However, there is no direct relation between the two spectra, indicating that the vibration in the sample is not directly correlated to the vibration frequencies of the rotary valve itself.

\subsection{Spectro-temporal analysis}
\label{sec:spectrotemporal}
\begin{figure}[h!]
    \includegraphics[width=0.95\columnwidth]{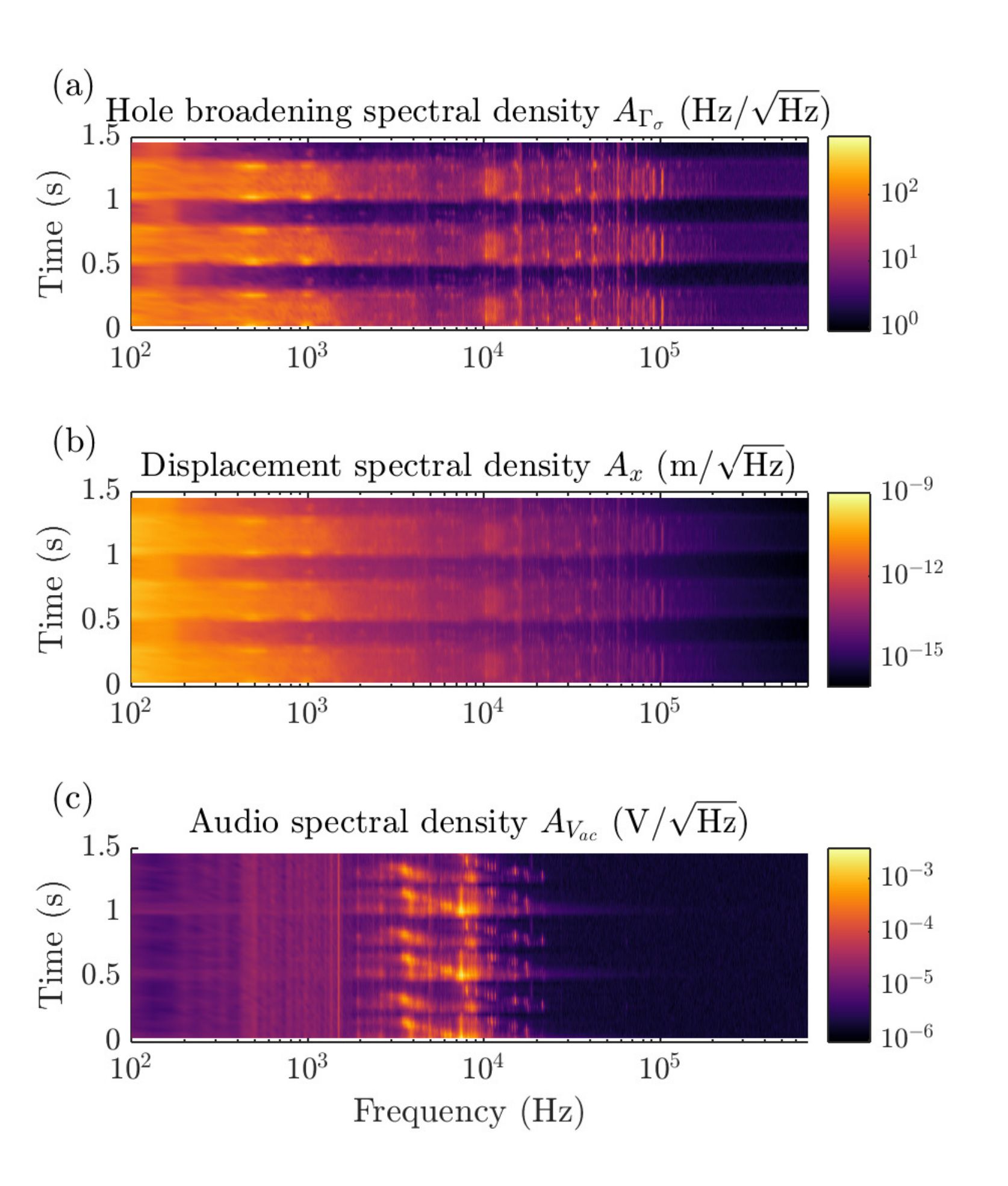}
    \caption{Amplitude spectral density spectrograms of (a) the spectral hole broadening, (b)  the corresponding displacement, and (c) the audio sound level on the rotary valve.}
    \label{fig:Spectres}
\end{figure}

We calculate the single-sided amplitude spectral density~\cite{bendatpiersol1971} of the hole broadening for each sequence cycle:
\begin{equation}
A_{\Gamma_\sigma}(f)=\sqrt{\frac 2T} |\tilde{\Gamma}_\sigma(f)|,
\end{equation}
where $\tilde{\Gamma}_\sigma(f)=\int_{-T/2}^{T/2} \Gamma_\sigma(t)e^{2i\pi ft} dt$ is the Fourier transform of the hole broadening $\Gamma_\sigma(t)$ over the burning pulse duration $T=30$~ms. $A_{\Gamma_\sigma}(f)$ is expressed in Hz\perroothz.
We calculate the hole broadening spectral density for each sequence cycle, and plot the result in the form of a spectrogram in Fig.~\ref{fig:Spectres}(a), averaged over 150 acquisitions. It should be noted that the integration time $T=30$~ms is characteristic of the population dynamics (longest state lifetime). In other words, the populations are roughly stationary during this period. This ensures that the transmission linearly follows the frequency shift produced by the vibrations. A longer integration time could induce a population change that would give a non-linear response.

In Eq.~\ref{eq:gamma} we gave the link between the hole broadening and the sample velocity. As explained in section~\ref{piezo_and_vib} we drop the absolute value in the equation. Displacement and velocity are linked by a derivation operation in the time-domain, which translates into $\tilde{\dot{x}}(f)=2i\pi f \tilde{x}(f)$ in the spectral domain. We thus obtain the displacement spectral density:
\begin{equation}
A_x(f)=\frac{V} {\kappa E 2\pi f}A_{\Gamma_\sigma}(f)
\label{eq:xtilde}
\end{equation}
that we display in Fig.~\ref{fig:Spectres}(b). Finally, Fig.~\ref{fig:Spectres}(c) shows the spectrogram of the audio signal spectral density $A_{V_{ac}}(f)$ measured on the rotary valve.

%The three spectrograms confirm the concomitance of acoustic noise and mechanical displacement of the cold finger.
The audio spectrogram clearly shows the rotary valve cycles at $2$~Hz, with the high-pitched chirp (around $10$~kHz) that can be heard twice per valve cycle, in agreement with previously reported cryocooler audio spectrograms~\cite{kalra2016vibration}. From the hole broadening and displacement spectrograms, we can define two uneven alternating steps in the rotary valve cycle that were already visible in Fig.~\ref{fig:TracesBrutes}: a noisier step, starting almost together with the loud rotary valve chirp and lasting $0.3~$s, followed by a quieter step lasting $0.2$~s. These two steps are not visible in the acoustic spectrogram, confirming that the vibration of the rotary valve is not  transmitted to the cold finger in a straightforward manner. By comparing the characteristic frequencies in the spectrograms of the spectral hole broadening and the audio signal, we could not observe any correlation neither, even by anticipating the frequency-doubling (second order harmonics) due the crude removal of the absolute value in Eq.~\ref{eq:gamma}.

\begin{figure}[h!]
\centering
    \includegraphics[width=0.95\columnwidth]{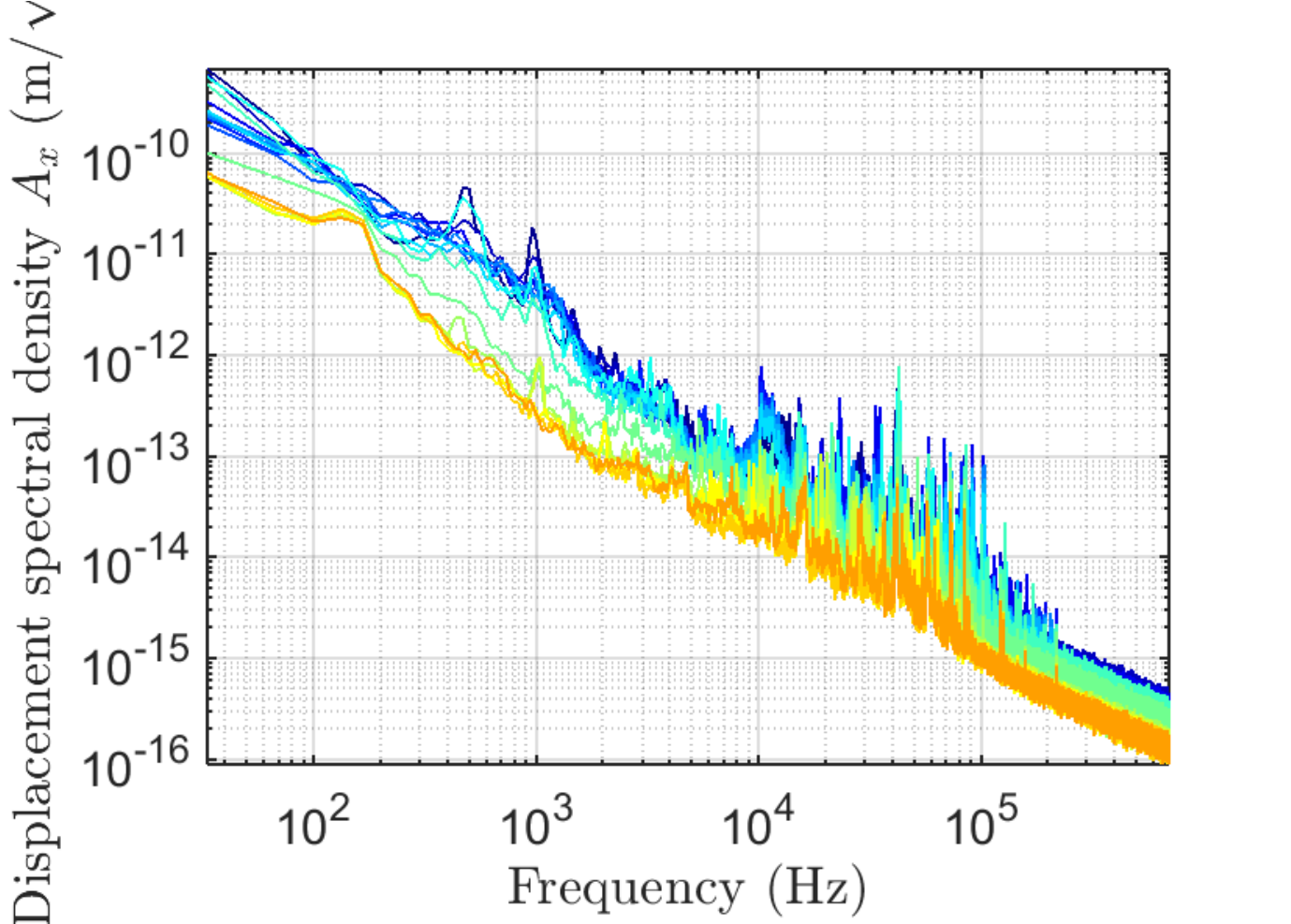}
    \caption{Cold finger displacement spectral density derived from the absorption of a spectral hole in a Tm-doped crystal at $2.9$~K. The lines are colored according to their position in the rotary valve cycle, from blue at the beginning to orange at the end.}
    \label{fig:SpectresDispl}
\end{figure}

\subsection{Vibration spectra}
In Fig.~\ref{fig:SpectresDispl}, we plot the displacement spectral density  as a multiplot with the same color code as in Fig~\ref{fig:SpectresHB}.
At $1$~kHz, $A_{x}(f)$ is equal to  $2\cdot10^{-11}$~m\perroothz in the noisy phase, and $2\cdot10^{-13}$~m\perroothz in the quiet phase. This is two to three orders of magnitude lower than the values measured in other pulse-tube cryocoolers with no vibration isolation (between $10^{-8}$~m\perroothz with optical detection~\cite{tomaru2004vibration} and $5\cdot 10^{-10}$~m\perroothz with accelerometric detection~\cite{majorana2006vibration}). This discrepancy could be explained by an inaccurate estimation of the displacement, or a lower vibration level of our low-power pulse-tube cooler compared to previously measured instruments. We discuss these in turn.
%However, the frequency dependence between $1/f$ and $1/f^2$
There is an inaccuracy in the quantitative estimation of the vibration amplitude, because of the  simplistic model used to relate the crystal absorption to the local strain, the somewhat empirical determination of the piezospectroscopic coefficient $\kappa$ (by taking a typical averaged value of the measured shifts in \ref{shb_tmyag}), and the assumption of only positive strain.

It should be also noted that the originality of our method makes the comparison with other optical vibration measurements difficult. This is a general issue when comparing different cryocoolers characterized with different techniques. As an example, our pulse-tube has a low input power (2kW) as compared to other vibrations studies \cite{tomaru2004vibration, majorana2006vibration} (typically 7kW). This makes direct comparisons generally challenging.

More fundamentally, the piezospectroscopic effect is sensitive to the local stress applied on the crystal attached to the second stage. Other optical methods intrinsically measure the relative position of the cold plate with respect to a reference point on the pulse-tube or the optical table base plane. There is a priori no good or bad method. The appropriateness depends on the measurement of interest driven by the application. Different techniques should be seen as complementary to give a more complete characterization of the vibrations. In that sense, our measurements could be more quantitative if connected at lower frequencies to another type of measurement ({\it ie} with 3-axis accelerometers).

%considerations on the vibration detection method
Due to its intrinsic sensitivity to velocity, our method's sensitivity to displacement increases with the frequency. It operates between $100$~Hz and $1$~MHz. The lower value is imposed by the spectral hole lifetime ($10$~ms for Tm:YAG~\cite[chap.7]{liu2006spectroscopic}), whereas the upper value is a purely technical limit since it corresponds to the Nyquist frequency, {\it ie} $1$~MHz.

%\subsection{Correlation with the rotary valve acoustic recording}

\section{Conclusion}
In this paper, we presented a novel optical method based on the piezospectroscopic effect in a rare-earth ion-doped crystal for the detection of high frequency ($>100$~Hz) vibration in a cold environment. Although not fully quantitative, this method is multidirectional and contact-less. Its sensitivity increases with frequency and has no fundamental upper frequency limit.

For our demonstration, we have used a cryocooler with optical windows. Nonetheless, further integration of the setup (primarily the crystal and the detector) is totally possible to form a compact single-component sensor with a fibered input feed-through (probe laser) and electric readout port (photodetector output). This integration step would allow to characterize a much wider range of cryocoolers without optical access. If the contact-less character of the method should be preserved, the fiber port can be attached to the cryocooler first stage and the crystal to the second stage (two-component sensor)

\section{Acknowledgments}
We are grateful to E. Olivieri and C. Marrache-Kikuchi for stimulating discussions. We thank also J. Paris (My Cryo Firm), J. Falter (TransMIT) and P. Pariset (CNRS-LAC) for technical assistance on the pulse-tube maintenance and design.

This work was supported by ITMO Cancer: AVIESAN (National Alliance for Life Sciences \& Health) within the framework of the Cancer Plan, by the Investissements d'Avenir du LabEx PALM ExciMol and OptoRF-Er (ANR-10-LABX-0039-PALM), by the DIM Nano-K project RECTUS and by the CMDO+ network (Mission pour l'Interdisciplinarit\'e du CNRS).

\appendix
\section{Discarding the Doppler effect}
In this paper, we have identified the piezospectroscopic effect as the principal source of coupling between vibrations and the optical absorption. As we discussed, rare-earth transitions are quite sensitive to strain because of their exceptionally narrow linewidth. This atomic-like feature for a solid impurity may give the impression that the Doppler shift induced by the vibrations could be a source of perturbation. We take the opportunity to show that the Doppler shift or broadening is much weaker than the piezospectroscopic effect.

The Doppler effect that would induce a similar broadening that the one discussed in \ref{piezo_and_vib}. The latter is also proportional to the crystal velocity. It is given by
\begin{equation}
\Gamma_D^\mathrm{[Hz]} \left(t\right) = \frac{1}{\lambda}|\dot{x}\left(t\right)| = 1.3\times10^6 \, |\dot{x}^\mathrm{[m/s]}\left(t\right)|
\end{equation}
where $\lambda=793$~nm is the transition wavelength for our crystal. A direct comparison with Eq.\eqref{eq:gamma} is then possible. The Doppler effect is three orders of magnitude weaker than the piezospectroscopic shift and can be fully neglected in first approach.
%\section{Perspective: Fabry-Perot vibrational measurement}

\bibliography{vibrations_bib}

%merlin.mbs aipnum4-1.bst 2010-07-25 4.21a (PWD, AO, DPC) hacked
%Control: key (0)
%Control: author (8) initials jnrlst
%Control: editor formatted (1) identically to author
%Control: production of article title (0) allowed
%Control: page (1) range
%Control: year (1) truncated
%Control: production of eprint (0) enabled
\begin{thebibliography}{50}%
\makeatletter
\providecommand \@ifxundefined [1]{%
 \@ifx{#1\undefined}
}%
\providecommand \@ifnum [1]{%
 \ifnum #1\expandafter \@firstoftwo
 \else \expandafter \@secondoftwo
 \fi
}%
\providecommand \@ifx [1]{%
 \ifx #1\expandafter \@firstoftwo
 \else \expandafter \@secondoftwo
 \fi
}%
\providecommand \natexlab [1]{#1}%
\providecommand \enquote  [1]{``#1''}%
\providecommand \bibnamefont  [1]{#1}%
\providecommand \bibfnamefont [1]{#1}%
\providecommand \citenamefont [1]{#1}%
\providecommand \href@noop [0]{\@secondoftwo}%
\providecommand \href [0]{\begingroup \@sanitize@url \@href}%
\providecommand \@href[1]{\@@startlink{#1}\@@href}%
\providecommand \@@href[1]{\endgroup#1\@@endlink}%
\providecommand \@sanitize@url [0]{\catcode `\\12\catcode `\$12\catcode
  `\&12\catcode `\#12\catcode `\^12\catcode `\_12\catcode `\%12\relax}%
\providecommand \@@startlink[1]{}%
\providecommand \@@endlink[0]{}%
\providecommand \url  [0]{\begingroup\@sanitize@url \@url }%
\providecommand \@url [1]{\endgroup\@href {#1}{\urlprefix }}%
\providecommand \urlprefix  [0]{URL }%
\providecommand \Eprint [0]{\href }%
\providecommand \doibase [0]{http://dx.doi.org/}%
\providecommand \selectlanguage [0]{\@gobble}%
\providecommand \bibinfo  [0]{\@secondoftwo}%
\providecommand \bibfield  [0]{\@secondoftwo}%
\providecommand \translation [1]{[#1]}%
\providecommand \BibitemOpen [0]{}%
\providecommand \bibitemStop [0]{}%
\providecommand \bibitemNoStop [0]{.\EOS\space}%
\providecommand \EOS [0]{\spacefactor3000\relax}%
\providecommand \BibitemShut  [1]{\csname bibitem#1\endcsname}%
\let\auto@bib@innerbib\@empty
%</preamble>
\bibitem [{\citenamefont {Oh}, \citenamefont {Lee},\ and\ \citenamefont
  {Jo}(2013)}]{oh2013passive}%
  \BibitemOpen
  \bibfield  {author} {\bibinfo {author} {\bibfnamefont {H.}~\bibnamefont
  {Oh}}, \bibinfo {author} {\bibfnamefont {K.}~\bibnamefont {Lee}}, \ and\
  \bibinfo {author} {\bibfnamefont {M.}~\bibnamefont {Jo}},\ }\bibfield
  {title} {\enquote {\bibinfo {title} {A passive launch and on-orbit vibration
  isolation system for the spaceborne cryocooler},}\ }\href {\doibase
  https://doi.org/10.1016/j.ast.2012.11.013} {\bibfield  {journal} {\bibinfo
  {journal} {Aerospace Science and Technology}\ }\textbf {\bibinfo {volume}
  {28}},\ \bibinfo {pages} {324--331} (\bibinfo {year} {2013})}\BibitemShut
  {NoStop}%
\bibitem [{\citenamefont {Radebaugh}(2009)}]{radebaugh2009cryocoolers}%
  \BibitemOpen
  \bibfield  {author} {\bibinfo {author} {\bibfnamefont {R.}~\bibnamefont
  {Radebaugh}},\ }\bibfield  {title} {\enquote {\bibinfo {title} {Cryocoolers:
  the state of the art and recent developments},}\ }\href@noop {} {\bibfield
  {journal} {\bibinfo  {journal} {Journal of Physics: Condensed Matter}\
  }\textbf {\bibinfo {volume} {21}},\ \bibinfo {pages} {164219} (\bibinfo
  {year} {2009})}\BibitemShut {NoStop}%
\bibitem [{\citenamefont {Tomaru}\ \emph {et~al.}(2004)\citenamefont {Tomaru},
  \citenamefont {Suzuki}, \citenamefont {Haruyama}, \citenamefont {Shintomi},
  \citenamefont {Yamamoto}, \citenamefont {Koyama},\ and\ \citenamefont
  {Li}}]{tomaru2004vibration}%
  \BibitemOpen
  \bibfield  {author} {\bibinfo {author} {\bibfnamefont {T.}~\bibnamefont
  {Tomaru}}, \bibinfo {author} {\bibfnamefont {T.}~\bibnamefont {Suzuki}},
  \bibinfo {author} {\bibfnamefont {T.}~\bibnamefont {Haruyama}}, \bibinfo
  {author} {\bibfnamefont {T.}~\bibnamefont {Shintomi}}, \bibinfo {author}
  {\bibfnamefont {A.}~\bibnamefont {Yamamoto}}, \bibinfo {author}
  {\bibfnamefont {T.}~\bibnamefont {Koyama}}, \ and\ \bibinfo {author}
  {\bibfnamefont {R.}~\bibnamefont {Li}},\ }\bibfield  {title} {\enquote
  {\bibinfo {title} {Vibration analysis of cryocoolers},}\ }\href@noop {}
  {\bibfield  {journal} {\bibinfo  {journal} {Cryogenics}\ }\textbf {\bibinfo
  {volume} {44}},\ \bibinfo {pages} {309--317} (\bibinfo {year}
  {2004})}\BibitemShut {NoStop}%
\bibitem [{\citenamefont {Tomaru}\ \emph {et~al.}(2005)\citenamefont {Tomaru},
  \citenamefont {Suzuki}, \citenamefont {Haruyama}, \citenamefont {Shintomi},
  \citenamefont {Sato}, \citenamefont {Yamamoto}, \citenamefont {Ikushima},
  \citenamefont {Li}, \citenamefont {Akutsu}, \citenamefont {Uchiyama},\ and\
  \citenamefont {Miyoki}}]{tomaru2005vibration}%
  \BibitemOpen
  \bibfield  {author} {\bibinfo {author} {\bibfnamefont {T.}~\bibnamefont
  {Tomaru}}, \bibinfo {author} {\bibfnamefont {T.}~\bibnamefont {Suzuki}},
  \bibinfo {author} {\bibfnamefont {T.}~\bibnamefont {Haruyama}}, \bibinfo
  {author} {\bibfnamefont {T.}~\bibnamefont {Shintomi}}, \bibinfo {author}
  {\bibfnamefont {N.}~\bibnamefont {Sato}}, \bibinfo {author} {\bibfnamefont
  {A.}~\bibnamefont {Yamamoto}}, \bibinfo {author} {\bibfnamefont
  {Y.}~\bibnamefont {Ikushima}}, \bibinfo {author} {\bibfnamefont
  {R.}~\bibnamefont {Li}}, \bibinfo {author} {\bibfnamefont {T.}~\bibnamefont
  {Akutsu}}, \bibinfo {author} {\bibfnamefont {T.}~\bibnamefont {Uchiyama}}, \
  and\ \bibinfo {author} {\bibfnamefont {S.}~\bibnamefont {Miyoki}},\
  }\bibfield  {title} {\enquote {\bibinfo {title} {Vibration-free pulse tube
  cryocooler system for gravitational wave detectors, part i:
  Vibration-reduction method and measurement},}\ }in\ \href@noop {} {\emph
  {\bibinfo {booktitle} {Cryocoolers 13}}},\ \bibinfo {editor} {edited by\
  \bibinfo {editor} {\bibfnamefont {R.~G.}\ \bibnamefont {Ross}}}\ (\bibinfo
  {publisher} {Springer US},\ \bibinfo {address} {Boston, MA},\ \bibinfo {year}
  {2005})\ pp.\ \bibinfo {pages} {695--702}\BibitemShut {NoStop}%
\bibitem [{\citenamefont {Grop}\ \emph {et~al.}(2010)\citenamefont {Grop},
  \citenamefont {Bourgeois}, \citenamefont {Bazin}, \citenamefont
  {Kersal{\'e}}, \citenamefont {Rubiola}, \citenamefont {Langham},
  \citenamefont {Oxborrow}, \citenamefont {Clapton}, \citenamefont {Walker},
  \citenamefont {De~Vicente} \emph {et~al.}}]{grop2010elisa}%
  \BibitemOpen
  \bibfield  {author} {\bibinfo {author} {\bibfnamefont {S.}~\bibnamefont
  {Grop}}, \bibinfo {author} {\bibfnamefont {P.}~\bibnamefont {Bourgeois}},
  \bibinfo {author} {\bibfnamefont {N.}~\bibnamefont {Bazin}}, \bibinfo
  {author} {\bibfnamefont {Y.}~\bibnamefont {Kersal{\'e}}}, \bibinfo {author}
  {\bibfnamefont {E.}~\bibnamefont {Rubiola}}, \bibinfo {author} {\bibfnamefont
  {C.}~\bibnamefont {Langham}}, \bibinfo {author} {\bibfnamefont
  {M.}~\bibnamefont {Oxborrow}}, \bibinfo {author} {\bibfnamefont
  {D.}~\bibnamefont {Clapton}}, \bibinfo {author} {\bibfnamefont
  {S.}~\bibnamefont {Walker}}, \bibinfo {author} {\bibfnamefont
  {J.}~\bibnamefont {De~Vicente}},  \emph {et~al.},\ }\bibfield  {title}
  {\enquote {\bibinfo {title} {{ELISA}: A cryocooled $10$ {GHz} oscillator with
  $10^{-15}$ frequency stability},}\ }\href {\doibase
  https://doi.org/10.1063/1.3290631} {\bibfield  {journal} {\bibinfo  {journal}
  {Review of Scientific Instruments}\ }\textbf {\bibinfo {volume} {81}},\
  \bibinfo {pages} {025102} (\bibinfo {year} {2010})}\BibitemShut {NoStop}%
\bibitem [{\citenamefont {Caparrelli}\ \emph {et~al.}(2006)\citenamefont
  {Caparrelli}, \citenamefont {Majorana}, \citenamefont {Moscatelli},
  \citenamefont {Pascucci}, \citenamefont {Perciballi}, \citenamefont {Puppo},
  \citenamefont {Rapagnani},\ and\ \citenamefont
  {Ricci}}]{caparrelli2006vibration}%
  \BibitemOpen
  \bibfield  {author} {\bibinfo {author} {\bibfnamefont {S.}~\bibnamefont
  {Caparrelli}}, \bibinfo {author} {\bibfnamefont {E.}~\bibnamefont
  {Majorana}}, \bibinfo {author} {\bibfnamefont {V.}~\bibnamefont
  {Moscatelli}}, \bibinfo {author} {\bibfnamefont {E.}~\bibnamefont
  {Pascucci}}, \bibinfo {author} {\bibfnamefont {M.}~\bibnamefont
  {Perciballi}}, \bibinfo {author} {\bibfnamefont {P.}~\bibnamefont {Puppo}},
  \bibinfo {author} {\bibfnamefont {P.}~\bibnamefont {Rapagnani}}, \ and\
  \bibinfo {author} {\bibfnamefont {F.}~\bibnamefont {Ricci}},\ }\bibfield
  {title} {\enquote {\bibinfo {title} {Vibration-free cryostat for low-noise
  applications of a pulse tube cryocooler},}\ }\href@noop {} {\bibfield
  {journal} {\bibinfo  {journal} {Review of Scientific Instruments}\ }\textbf
  {\bibinfo {volume} {77}},\ \bibinfo {pages} {095102} (\bibinfo {year}
  {2006})}\BibitemShut {NoStop}%
\bibitem [{\citenamefont {Evans}\ \emph {et~al.}(2008)\citenamefont {Evans},
  \citenamefont {Down}, \citenamefont {Keeping}, \citenamefont {Kirichek},\
  and\ \citenamefont {Bowden}}]{evans2008cryogen}%
  \BibitemOpen
  \bibfield  {author} {\bibinfo {author} {\bibfnamefont {B.}~\bibnamefont
  {Evans}}, \bibinfo {author} {\bibfnamefont {R.}~\bibnamefont {Down}},
  \bibinfo {author} {\bibfnamefont {J.}~\bibnamefont {Keeping}}, \bibinfo
  {author} {\bibfnamefont {O.}~\bibnamefont {Kirichek}}, \ and\ \bibinfo
  {author} {\bibfnamefont {Z.}~\bibnamefont {Bowden}},\ }\bibfield  {title}
  {\enquote {\bibinfo {title} {Cryogen-free low temperature sample environment
  for neutron scattering based on pulse tube refrigeration},}\ }\href@noop {}
  {\bibfield  {journal} {\bibinfo  {journal} {Measurement Science and
  Technology}\ }\textbf {\bibinfo {volume} {19}},\ \bibinfo {pages} {034018}
  (\bibinfo {year} {2008})}\BibitemShut {NoStop}%
\bibitem [{\citenamefont {Wang}\ and\ \citenamefont
  {Hartnett}(2010)}]{wang2010vibration}%
  \BibitemOpen
  \bibfield  {author} {\bibinfo {author} {\bibfnamefont {C.}~\bibnamefont
  {Wang}}\ and\ \bibinfo {author} {\bibfnamefont {J.~G.}\ \bibnamefont
  {Hartnett}},\ }\bibfield  {title} {\enquote {\bibinfo {title} {A vibration
  free cryostat using pulse tube cryocooler},}\ }\href@noop {} {\bibfield
  {journal} {\bibinfo  {journal} {Cryogenics}\ }\textbf {\bibinfo {volume}
  {50}},\ \bibinfo {pages} {336--341} (\bibinfo {year} {2010})}\BibitemShut
  {NoStop}%
\bibitem [{\citenamefont {Hackley}\ \emph {et~al.}(2014)\citenamefont
  {Hackley}, \citenamefont {Kislitsyn}, \citenamefont {Beaman}, \citenamefont
  {Ulrich},\ and\ \citenamefont {Nazin}}]{stm}%
  \BibitemOpen
  \bibfield  {author} {\bibinfo {author} {\bibfnamefont {J.}~\bibnamefont
  {Hackley}}, \bibinfo {author} {\bibfnamefont {D.}~\bibnamefont {Kislitsyn}},
  \bibinfo {author} {\bibfnamefont {D.}~\bibnamefont {Beaman}}, \bibinfo
  {author} {\bibfnamefont {S.}~\bibnamefont {Ulrich}}, \ and\ \bibinfo {author}
  {\bibfnamefont {G.}~\bibnamefont {Nazin}},\ }\bibfield  {title} {\enquote
  {\bibinfo {title} {High-stability cryogenic scanning tunneling microscope
  based on a closed-cycle cryostat},}\ }\href {\doibase 10.1063/1.4897139}
  {\bibfield  {journal} {\bibinfo  {journal} {Review of Scientific
  Instruments}\ }\textbf {\bibinfo {volume} {85}},\ \bibinfo {pages} {103704}
  (\bibinfo {year} {2014})}\BibitemShut {NoStop}%
\bibitem [{\citenamefont {Quacquarelli}\ \emph {et~al.}(2015)\citenamefont
  {Quacquarelli}, \citenamefont {Puebla}, \citenamefont {Scheler},
  \citenamefont {Andres}, \citenamefont {B{\"o}defeld}, \citenamefont {Sipos},
  \citenamefont {Dal~Savio}, \citenamefont {Bauer}, \citenamefont {Pfleiderer},
  \citenamefont {Erb} \emph {et~al.}}]{quacquarelli2015scanning}%
  \BibitemOpen
  \bibfield  {author} {\bibinfo {author} {\bibfnamefont {F.}~\bibnamefont
  {Quacquarelli}}, \bibinfo {author} {\bibfnamefont {J.}~\bibnamefont
  {Puebla}}, \bibinfo {author} {\bibfnamefont {T.}~\bibnamefont {Scheler}},
  \bibinfo {author} {\bibfnamefont {D.}~\bibnamefont {Andres}}, \bibinfo
  {author} {\bibfnamefont {C.}~\bibnamefont {B{\"o}defeld}}, \bibinfo {author}
  {\bibfnamefont {B.}~\bibnamefont {Sipos}}, \bibinfo {author} {\bibfnamefont
  {C.}~\bibnamefont {Dal~Savio}}, \bibinfo {author} {\bibfnamefont
  {A.}~\bibnamefont {Bauer}}, \bibinfo {author} {\bibfnamefont
  {C.}~\bibnamefont {Pfleiderer}}, \bibinfo {author} {\bibfnamefont
  {A.}~\bibnamefont {Erb}},  \emph {et~al.},\ }\bibfield  {title} {\enquote
  {\bibinfo {title} {Scanning probe microscopy in an ultra-low vibration
  closed-cycle cryostat: Skyrmion lattice detection and tuning fork
  implementation},}\ }\href@noop {} {\bibfield  {journal} {\bibinfo  {journal}
  {Microscopy Today}\ }\textbf {\bibinfo {volume} {23}},\ \bibinfo {pages}
  {12--17} (\bibinfo {year} {2015})}\BibitemShut {NoStop}%
\bibitem [{\citenamefont {Maisonobe}\ \emph {et~al.}(2018)\citenamefont
  {Maisonobe}, \citenamefont {Billard}, \citenamefont {De~Jesus}, \citenamefont
  {Juillard}, \citenamefont {Misiak}, \citenamefont {Olivieri}, \citenamefont
  {Sayah},\ and\ \citenamefont {Vagneron}}]{maisonobe2018vibration}%
  \BibitemOpen
  \bibfield  {author} {\bibinfo {author} {\bibfnamefont {R.}~\bibnamefont
  {Maisonobe}}, \bibinfo {author} {\bibfnamefont {J.}~\bibnamefont {Billard}},
  \bibinfo {author} {\bibfnamefont {M.}~\bibnamefont {De~Jesus}}, \bibinfo
  {author} {\bibfnamefont {A.}~\bibnamefont {Juillard}}, \bibinfo {author}
  {\bibfnamefont {D.}~\bibnamefont {Misiak}}, \bibinfo {author} {\bibfnamefont
  {E.}~\bibnamefont {Olivieri}}, \bibinfo {author} {\bibfnamefont
  {S.}~\bibnamefont {Sayah}}, \ and\ \bibinfo {author} {\bibfnamefont
  {L.}~\bibnamefont {Vagneron}},\ }\bibfield  {title} {\enquote {\bibinfo
  {title} {Vibration decoupling system for massive bolometers in dry
  cryostats},}\ }\href@noop {} {\bibfield  {journal} {\bibinfo  {journal}
  {arXiv preprint arXiv:1803.03463}\ } (\bibinfo {year} {2018})}\BibitemShut
  {NoStop}%
\bibitem [{\citenamefont {Chijioke}\ and\ \citenamefont
  {Lawall}(2010)}]{chijioke2010vibration}%
  \BibitemOpen
  \bibfield  {author} {\bibinfo {author} {\bibfnamefont {A.}~\bibnamefont
  {Chijioke}}\ and\ \bibinfo {author} {\bibfnamefont {J.}~\bibnamefont
  {Lawall}},\ }\bibfield  {title} {\enquote {\bibinfo {title} {Vibration
  spectrum of a pulse-tube cryostat from 1 {Hz} to 20 {kHz}},}\ }\href@noop {}
  {\bibfield  {journal} {\bibinfo  {journal} {Cryogenics}\ }\textbf {\bibinfo
  {volume} {50}},\ \bibinfo {pages} {266--270} (\bibinfo {year}
  {2010})}\BibitemShut {NoStop}%
\bibitem [{\citenamefont {Riabzev}\ \emph {et~al.}(2009)\citenamefont
  {Riabzev}, \citenamefont {Veprik}, \citenamefont {Vilenchik},\ and\
  \citenamefont {Pundak}}]{riabzev2009vibration}%
  \BibitemOpen
  \bibfield  {author} {\bibinfo {author} {\bibfnamefont {S.}~\bibnamefont
  {Riabzev}}, \bibinfo {author} {\bibfnamefont {A.}~\bibnamefont {Veprik}},
  \bibinfo {author} {\bibfnamefont {H.}~\bibnamefont {Vilenchik}}, \ and\
  \bibinfo {author} {\bibfnamefont {N.}~\bibnamefont {Pundak}},\ }\bibfield
  {title} {\enquote {\bibinfo {title} {Vibration generation in a pulse tube
  refrigerator},}\ }\href@noop {} {\bibfield  {journal} {\bibinfo  {journal}
  {Cryogenics}\ }\textbf {\bibinfo {volume} {49}},\ \bibinfo {pages} {1--6}
  (\bibinfo {year} {2009})}\BibitemShut {NoStop}%
\bibitem [{\citenamefont {Schwab}\ and\ \citenamefont
  {Roukes}(2005)}]{schwab2005putting}%
  \BibitemOpen
  \bibfield  {author} {\bibinfo {author} {\bibfnamefont {K.}~\bibnamefont
  {Schwab}}\ and\ \bibinfo {author} {\bibfnamefont {M.}~\bibnamefont
  {Roukes}},\ }\bibfield  {title} {\enquote {\bibinfo {title} {Putting
  mechanics into quantum mechanics},}\ }\href@noop {} {\bibfield  {journal}
  {\bibinfo  {journal} {Physics Today}\ }\textbf {\bibinfo {volume} {58}},\
  \bibinfo {pages} {36--42} (\bibinfo {year} {2005})}\BibitemShut {NoStop}%
\bibitem [{\citenamefont {Majorana}\ \emph {et~al.}(2006)\citenamefont
  {Majorana}, \citenamefont {Perciballi}, \citenamefont {Puppo}, \citenamefont
  {Papagnani},\ and\ \citenamefont {Ricci}}]{majorana2006vibration}%
  \BibitemOpen
  \bibfield  {author} {\bibinfo {author} {\bibfnamefont {E.}~\bibnamefont
  {Majorana}}, \bibinfo {author} {\bibfnamefont {M.}~\bibnamefont
  {Perciballi}}, \bibinfo {author} {\bibfnamefont {P.}~\bibnamefont {Puppo}},
  \bibinfo {author} {\bibfnamefont {P.}~\bibnamefont {Papagnani}}, \ and\
  \bibinfo {author} {\bibfnamefont {F.}~\bibnamefont {Ricci}},\ }\bibfield
  {title} {\enquote {\bibinfo {title} {Vibration free cryostat for cooling
  suspended mirrors},}\ }in\ \href@noop {} {\emph {\bibinfo {booktitle}
  {Journal of Physics: Conference Series}}},\ Vol.~\bibinfo {volume} {32}\
  (\bibinfo {organization} {IOP Publishing},\ \bibinfo {year} {2006})\ p.\
  \bibinfo {pages} {374}\BibitemShut {NoStop}%
\bibitem [{\citenamefont {Mauritsen}\ \emph {et~al.}(2009)\citenamefont
  {Mauritsen}, \citenamefont {Snow}, \citenamefont {Woidtke}, \citenamefont
  {Chase},\ and\ \citenamefont {Henslee}}]{mauritsen2009low}%
  \BibitemOpen
  \bibfield  {author} {\bibinfo {author} {\bibfnamefont {L.}~\bibnamefont
  {Mauritsen}}, \bibinfo {author} {\bibfnamefont {D.}~\bibnamefont {Snow}},
  \bibinfo {author} {\bibfnamefont {A.}~\bibnamefont {Woidtke}}, \bibinfo
  {author} {\bibfnamefont {M.}~\bibnamefont {Chase}}, \ and\ \bibinfo {author}
  {\bibfnamefont {I.}~\bibnamefont {Henslee}},\ }\bibfield  {title} {\enquote
  {\bibinfo {title} {Low vibration, low thermal fluctuation system for pulse
  tube and gifford-mcmahon cryocoolers},}\ }\href@noop {} {\bibfield  {journal}
  {\bibinfo  {journal} {Cryocoolers}\ }\textbf {\bibinfo {volume} {15}},\
  \bibinfo {pages} {581--585} (\bibinfo {year} {2009})}\BibitemShut {NoStop}%
\bibitem [{\citenamefont {He}\ and\ \citenamefont
  {Clarke}(1995)}]{he1995determination}%
  \BibitemOpen
  \bibfield  {author} {\bibinfo {author} {\bibfnamefont {J.}~\bibnamefont
  {He}}\ and\ \bibinfo {author} {\bibfnamefont {D.}~\bibnamefont {Clarke}},\
  }\bibfield  {title} {\enquote {\bibinfo {title} {Determination of the
  piezospectroscopic coefficients for chromium-doped sapphire},}\ }\href@noop
  {} {\bibfield  {journal} {\bibinfo  {journal} {Journal of the American
  Ceramic Society}\ }\textbf {\bibinfo {volume} {78}},\ \bibinfo {pages}
  {1347--1353} (\bibinfo {year} {1995})}\BibitemShut {NoStop}%
\bibitem [{\citenamefont {Christensen}\ \emph {et~al.}(1996)\citenamefont
  {Christensen}, \citenamefont {Lipkin}, \citenamefont {Clarke},\ and\
  \citenamefont {Murphy}}]{christensen1996nondestructive}%
  \BibitemOpen
  \bibfield  {author} {\bibinfo {author} {\bibfnamefont {R.}~\bibnamefont
  {Christensen}}, \bibinfo {author} {\bibfnamefont {D.}~\bibnamefont {Lipkin}},
  \bibinfo {author} {\bibfnamefont {D.~R.}\ \bibnamefont {Clarke}}, \ and\
  \bibinfo {author} {\bibfnamefont {K.}~\bibnamefont {Murphy}},\ }\bibfield
  {title} {\enquote {\bibinfo {title} {Nondestructive evaluation of the
  oxidation stresses through thermal barrier coatings using {Cr3+}
  piezospectroscopy},}\ }\href@noop {} {\bibfield  {journal} {\bibinfo
  {journal} {Applied Physics Letters}\ }\textbf {\bibinfo {volume} {69}},\
  \bibinfo {pages} {3754--3756} (\bibinfo {year} {1996})}\BibitemShut {NoStop}%
\bibitem [{\citenamefont {Schlichting}\ \emph {et~al.}(2000)\citenamefont
  {Schlichting}, \citenamefont {Vaidyanathan}, \citenamefont {Sohn},
  \citenamefont {Jordan}, \citenamefont {Gell},\ and\ \citenamefont
  {Padture}}]{schlichting2000application}%
  \BibitemOpen
  \bibfield  {author} {\bibinfo {author} {\bibfnamefont {K.}~\bibnamefont
  {Schlichting}}, \bibinfo {author} {\bibfnamefont {K.}~\bibnamefont
  {Vaidyanathan}}, \bibinfo {author} {\bibfnamefont {Y.}~\bibnamefont {Sohn}},
  \bibinfo {author} {\bibfnamefont {E.}~\bibnamefont {Jordan}}, \bibinfo
  {author} {\bibfnamefont {M.}~\bibnamefont {Gell}}, \ and\ \bibinfo {author}
  {\bibfnamefont {N.}~\bibnamefont {Padture}},\ }\bibfield  {title} {\enquote
  {\bibinfo {title} {Application of {Cr$^{3+}$} photoluminescence
  piezo-spectroscopy to plasma-sprayed thermal barrier coatings for residual
  stress measurement},}\ }\href@noop {} {\bibfield  {journal} {\bibinfo
  {journal} {Materials Science and Engineering: A}\ }\textbf {\bibinfo {volume}
  {291}},\ \bibinfo {pages} {68--77} (\bibinfo {year} {2000})}\BibitemShut
  {NoStop}%
\bibitem [{\citenamefont {Rajendran}\ \emph {et~al.}(2017)\citenamefont
  {Rajendran}, \citenamefont {Zobrist}, \citenamefont {Sushkov}, \citenamefont
  {Walsworth},\ and\ \citenamefont {Lukin}}]{rajendran2017wimp}%
  \BibitemOpen
  \bibfield  {author} {\bibinfo {author} {\bibfnamefont {S.}~\bibnamefont
  {Rajendran}}, \bibinfo {author} {\bibfnamefont {N.}~\bibnamefont {Zobrist}},
  \bibinfo {author} {\bibfnamefont {A.~O.}\ \bibnamefont {Sushkov}}, \bibinfo
  {author} {\bibfnamefont {R.}~\bibnamefont {Walsworth}}, \ and\ \bibinfo
  {author} {\bibfnamefont {M.}~\bibnamefont {Lukin}},\ }\bibfield  {title}
  {\enquote {\bibinfo {title} {A method for directional detection of dark
  matter using spectroscopy of crystal defects},}\ }\href {\doibase
  https://doi.org/10.1103/PhysRevD.96.035009} {\bibfield  {journal} {\bibinfo
  {journal} {Physical Review D}\ }\textbf {\bibinfo {volume} {96}},\ \bibinfo
  {pages} {035009} (\bibinfo {year} {2017})}\BibitemShut {NoStop}%
\bibitem [{Note1()}]{Note1}%
  \BibitemOpen
  \bibinfo {note} {As an order of magnitude, $400$~GHz corresponds to $1$~nm in
  the near IR}\BibitemShut {NoStop}%
\bibitem [{\citenamefont {Liu}\ and\ \citenamefont
  {Jacquier}(2006)}]{liu2006spectroscopic}%
  \BibitemOpen
  \bibfield  {author} {\bibinfo {author} {\bibfnamefont {G.}~\bibnamefont
  {Liu}}\ and\ \bibinfo {author} {\bibfnamefont {B.}~\bibnamefont {Jacquier}},\
  }\href@noop {} {\emph {\bibinfo {title} {Spectroscopic properties of rare
  earths in optical materials}}},\ Vol.~\bibinfo {volume} {83}\ (\bibinfo
  {publisher} {Springer Science \& Business Media},\ \bibinfo {year}
  {2006})\BibitemShut {NoStop}%
\bibitem [{\citenamefont {Merkel}\ \emph {et~al.}(2004)\citenamefont {Merkel},
  \citenamefont {Mohan}, \citenamefont {Cole}, \citenamefont {Chang},
  \citenamefont {Olson},\ and\ \citenamefont {Babbitt}}]{merkel2004multi}%
  \BibitemOpen
  \bibfield  {author} {\bibinfo {author} {\bibfnamefont {K.}~\bibnamefont
  {Merkel}}, \bibinfo {author} {\bibfnamefont {R.~K.}\ \bibnamefont {Mohan}},
  \bibinfo {author} {\bibfnamefont {Z.}~\bibnamefont {Cole}}, \bibinfo {author}
  {\bibfnamefont {T.}~\bibnamefont {Chang}}, \bibinfo {author} {\bibfnamefont
  {A.}~\bibnamefont {Olson}}, \ and\ \bibinfo {author} {\bibfnamefont
  {W.}~\bibnamefont {Babbitt}},\ }\bibfield  {title} {\enquote {\bibinfo
  {title} {{Multi-gigahertz radar range processing of baseband and {RF} carrier
  modulated signals in {Tm:YAG}}},}\ }\href@noop {} {\bibfield  {journal}
  {\bibinfo  {journal} {Journal of Luminescence}\ }\textbf {\bibinfo {volume}
  {107}},\ \bibinfo {pages} {62--74} (\bibinfo {year} {2004})}\BibitemShut
  {NoStop}%
\bibitem [{\citenamefont {Gorju}\ \emph {et~al.}(2005)\citenamefont {Gorju},
  \citenamefont {Crozatier}, \citenamefont {Lorger{\'e}}, \citenamefont
  {Le~Gou{\"e}t},\ and\ \citenamefont {Bretenaker}}]{gorju200510}%
  \BibitemOpen
  \bibfield  {author} {\bibinfo {author} {\bibfnamefont {G.}~\bibnamefont
  {Gorju}}, \bibinfo {author} {\bibfnamefont {V.}~\bibnamefont {Crozatier}},
  \bibinfo {author} {\bibfnamefont {I.}~\bibnamefont {Lorger{\'e}}}, \bibinfo
  {author} {\bibfnamefont {J.-L.}\ \bibnamefont {Le~Gou{\"e}t}}, \ and\
  \bibinfo {author} {\bibfnamefont {F.}~\bibnamefont {Bretenaker}},\ }\bibfield
   {title} {\enquote {\bibinfo {title} {{{10-GHz} bandwidth {RF} spectral
  analyzer with {MHz} resolution based on spectral hole burning in
  {Tm$^{3+}$:YAG}}},}\ }\href@noop {} {\bibfield  {journal} {\bibinfo
  {journal} {IEEE Photonics Technology Letters}\ }\textbf {\bibinfo {volume}
  {17}},\ \bibinfo {pages} {2385--2387} (\bibinfo {year} {2005})}\BibitemShut
  {NoStop}%
\bibitem [{\citenamefont {Li}\ \emph {et~al.}(2008)\citenamefont {Li},
  \citenamefont {Zhang}, \citenamefont {Kim}, \citenamefont {Wagner},
  \citenamefont {Hemmer},\ and\ \citenamefont {Wang}}]{li2008pulsed}%
  \BibitemOpen
  \bibfield  {author} {\bibinfo {author} {\bibfnamefont {Y.}~\bibnamefont
  {Li}}, \bibinfo {author} {\bibfnamefont {H.}~\bibnamefont {Zhang}}, \bibinfo
  {author} {\bibfnamefont {C.}~\bibnamefont {Kim}}, \bibinfo {author}
  {\bibfnamefont {K.}~\bibnamefont {Wagner}}, \bibinfo {author} {\bibfnamefont
  {P.}~\bibnamefont {Hemmer}}, \ and\ \bibinfo {author} {\bibfnamefont
  {L.}~\bibnamefont {Wang}},\ }\bibfield  {title} {\enquote {\bibinfo {title}
  {Pulsed ultrasound-modulated optical tomography using spectral-hole burning
  as a narrowband spectral filter},}\ }\href@noop {} {\bibfield  {journal}
  {\bibinfo  {journal} {Applied Physics Letters}\ }\textbf {\bibinfo {volume}
  {93}},\ \bibinfo {pages} {011111} (\bibinfo {year} {2008})}\BibitemShut
  {NoStop}%
\bibitem [{\citenamefont {Heshami}\ \emph {et~al.}(2016)\citenamefont
  {Heshami}, \citenamefont {England}, \citenamefont {Humphreys}, \citenamefont
  {Bustard}, \citenamefont {Acosta}, \citenamefont {Nunn},\ and\ \citenamefont
  {Sussman}}]{heshami2016quantum}%
  \BibitemOpen
  \bibfield  {author} {\bibinfo {author} {\bibfnamefont {K.}~\bibnamefont
  {Heshami}}, \bibinfo {author} {\bibfnamefont {D.}~\bibnamefont {England}},
  \bibinfo {author} {\bibfnamefont {P.}~\bibnamefont {Humphreys}}, \bibinfo
  {author} {\bibfnamefont {P.}~\bibnamefont {Bustard}}, \bibinfo {author}
  {\bibfnamefont {V.}~\bibnamefont {Acosta}}, \bibinfo {author} {\bibfnamefont
  {J.}~\bibnamefont {Nunn}}, \ and\ \bibinfo {author} {\bibfnamefont
  {B.}~\bibnamefont {Sussman}},\ }\bibfield  {title} {\enquote {\bibinfo
  {title} {Quantum memories: emerging applications and recent advances},}\
  }\href@noop {} {\bibfield  {journal} {\bibinfo  {journal} {Journal of Modern
  Optics}\ }\textbf {\bibinfo {volume} {63}},\ \bibinfo {pages} {2005--2028}
  (\bibinfo {year} {2016})}\BibitemShut {NoStop}%
\bibitem [{\citenamefont {M\o{}lmer}, \citenamefont {Le~Coq},\ and\
  \citenamefont {Seidelin}(2016)}]{Seidelin}%
  \BibitemOpen
  \bibfield  {author} {\bibinfo {author} {\bibfnamefont {K.}~\bibnamefont
  {M\o{}lmer}}, \bibinfo {author} {\bibfnamefont {Y.}~\bibnamefont {Le~Coq}}, \
  and\ \bibinfo {author} {\bibfnamefont {S.}~\bibnamefont {Seidelin}},\
  }\bibfield  {title} {\enquote {\bibinfo {title} {Dispersive coupling between
  light and a rare-earth-ion--doped mechanical resonator},}\ }\href {\doibase
  10.1103/PhysRevA.94.053804} {\bibfield  {journal} {\bibinfo  {journal} {Phys.
  Rev. A}\ }\textbf {\bibinfo {volume} {94}},\ \bibinfo {pages} {053804}
  (\bibinfo {year} {2016})}\BibitemShut {NoStop}%
\bibitem [{\citenamefont {Thorpe}\ \emph {et~al.}(2011)\citenamefont {Thorpe},
  \citenamefont {Rippe}, \citenamefont {Fortier}, \citenamefont {Kirchner},\
  and\ \citenamefont {Rosenband}}]{thorpe2011frequency}%
  \BibitemOpen
  \bibfield  {author} {\bibinfo {author} {\bibfnamefont {M.}~\bibnamefont
  {Thorpe}}, \bibinfo {author} {\bibfnamefont {L.}~\bibnamefont {Rippe}},
  \bibinfo {author} {\bibfnamefont {T.}~\bibnamefont {Fortier}}, \bibinfo
  {author} {\bibfnamefont {M.}~\bibnamefont {Kirchner}}, \ and\ \bibinfo
  {author} {\bibfnamefont {T.}~\bibnamefont {Rosenband}},\ }\bibfield  {title}
  {\enquote {\bibinfo {title} {Frequency stabilization to 6$\times10^{-16}$ via
  spectral-hole burning},}\ }\href@noop {} {\bibfield  {journal} {\bibinfo
  {journal} {Nature Photonics}\ }\textbf {\bibinfo {volume} {5}},\ \bibinfo
  {pages} {688} (\bibinfo {year} {2011})}\BibitemShut {NoStop}%
\bibitem [{\citenamefont {Chen}, \citenamefont {Fernandez-Gonzalvo},\ and\
  \citenamefont {Longdell}(2016)}]{PhysRevB.94.075117}%
  \BibitemOpen
  \bibfield  {author} {\bibinfo {author} {\bibfnamefont {Y.}~\bibnamefont
  {Chen}}, \bibinfo {author} {\bibfnamefont {X.}~\bibnamefont
  {Fernandez-Gonzalvo}}, \ and\ \bibinfo {author} {\bibfnamefont
  {J.}~\bibnamefont {Longdell}},\ }\bibfield  {title} {\enquote {\bibinfo
  {title} {Coupling erbium spins to a three-dimensional superconducting cavity
  at zero magnetic field},}\ }\href {\doibase 10.1103/PhysRevB.94.075117}
  {\bibfield  {journal} {\bibinfo  {journal} {Physical Review B}\ }\textbf
  {\bibinfo {volume} {94}},\ \bibinfo {pages} {075117} (\bibinfo {year}
  {2016})}\BibitemShut {NoStop}%
\bibitem [{\citenamefont {Thorpe}, \citenamefont {Leibrandt},\ and\
  \citenamefont {Rosenband}(2013)}]{thorpe2013shifts}%
  \BibitemOpen
  \bibfield  {author} {\bibinfo {author} {\bibfnamefont {M.}~\bibnamefont
  {Thorpe}}, \bibinfo {author} {\bibfnamefont {D.}~\bibnamefont {Leibrandt}}, \
  and\ \bibinfo {author} {\bibfnamefont {T.}~\bibnamefont {Rosenband}},\
  }\bibfield  {title} {\enquote {\bibinfo {title} {Shifts of optical frequency
  references based on spectral-hole burning in {Eu$^{3+}$:Y$_2$SiO$_5$}},}\
  }\href {http://stacks.iop.org/1367-2630/15/i=3/a=033006} {\bibfield
  {journal} {\bibinfo  {journal} {New Journal of Physics}\ }\textbf {\bibinfo
  {volume} {15}},\ \bibinfo {pages} {033006} (\bibinfo {year}
  {2013})}\BibitemShut {NoStop}%
\bibitem [{\citenamefont {Gobron}\ \emph {et~al.}(2017)\citenamefont {Gobron},
  \citenamefont {Jung}, \citenamefont {Galland}, \citenamefont {Predehl},
  \citenamefont {Le~Targat}, \citenamefont {Ferrier}, \citenamefont {Goldner},
  \citenamefont {Seidelin},\ and\ \citenamefont
  {Le~Coq}}]{gobron2017dispersive}%
  \BibitemOpen
  \bibfield  {author} {\bibinfo {author} {\bibfnamefont {O.}~\bibnamefont
  {Gobron}}, \bibinfo {author} {\bibfnamefont {K.}~\bibnamefont {Jung}},
  \bibinfo {author} {\bibfnamefont {N.}~\bibnamefont {Galland}}, \bibinfo
  {author} {\bibfnamefont {K.}~\bibnamefont {Predehl}}, \bibinfo {author}
  {\bibfnamefont {R.}~\bibnamefont {Le~Targat}}, \bibinfo {author}
  {\bibfnamefont {A.}~\bibnamefont {Ferrier}}, \bibinfo {author} {\bibfnamefont
  {P.}~\bibnamefont {Goldner}}, \bibinfo {author} {\bibfnamefont
  {S.}~\bibnamefont {Seidelin}}, \ and\ \bibinfo {author} {\bibfnamefont
  {Y.}~\bibnamefont {Le~Coq}},\ }\bibfield  {title} {\enquote {\bibinfo {title}
  {Dispersive heterodyne probing method for laser frequency stabilization based
  on spectral hole burning in rare-earth doped crystals},}\ }\href@noop {}
  {\bibfield  {journal} {\bibinfo  {journal} {Optics Express}\ }\textbf
  {\bibinfo {volume} {25}},\ \bibinfo {pages} {15539--15548} (\bibinfo {year}
  {2017})}\BibitemShut {NoStop}%
\bibitem [{\citenamefont {Kaplyanskii}(1964)}]{kaplyanskii1964noncubic}%
  \BibitemOpen
  \bibfield  {author} {\bibinfo {author} {\bibfnamefont {A.}~\bibnamefont
  {Kaplyanskii}},\ }\bibfield  {title} {\enquote {\bibinfo {title} {Noncubic
  centers in cubic crystals and their piezospectroscopic investigation},}\
  }\href@noop {} {\bibfield  {journal} {\bibinfo  {journal} {Optics and
  Spectroscopy}\ }\textbf {\bibinfo {volume} {16}},\ \bibinfo {pages} {329}
  (\bibinfo {year} {1964})}\BibitemShut {NoStop}%
\bibitem [{\citenamefont {Kaplianskii}\ and\ \citenamefont
  {Przhevuskii}(1965)}]{kaplianskii1965deformation}%
  \BibitemOpen
  \bibfield  {author} {\bibinfo {author} {\bibfnamefont {A.}~\bibnamefont
  {Kaplianskii}}\ and\ \bibinfo {author} {\bibfnamefont {A.}~\bibnamefont
  {Przhevuskii}},\ }\bibfield  {title} {\enquote {\bibinfo {title} {Deformation
  splitting and the evolution of spectral lines and the structure of the
  excited levels of {Eu}$^{2+}$ in crystals of alkaline-earth fluorides(plastic
  deformation splitting and evolution of spectral lines and excited level
  structure of europium ion in alkaline-earth fluoride crystals)},}\
  }\href@noop {} {\bibfield  {journal} {\bibinfo  {journal} {Optika I
  Spektroskopiia}\ }\textbf {\bibinfo {volume} {19}},\ \bibinfo {pages}
  {597--610} (\bibinfo {year} {1965})}\BibitemShut {NoStop}%
\bibitem [{\citenamefont {Bungenstock}, \citenamefont {Tr{\"o}ster},\ and\
  \citenamefont {Holzapfel}(2000)}]{bungenstock2000effect}%
  \BibitemOpen
  \bibfield  {author} {\bibinfo {author} {\bibfnamefont {C.}~\bibnamefont
  {Bungenstock}}, \bibinfo {author} {\bibfnamefont {T.}~\bibnamefont
  {Tr{\"o}ster}}, \ and\ \bibinfo {author} {\bibfnamefont {W.}~\bibnamefont
  {Holzapfel}},\ }\bibfield  {title} {\enquote {\bibinfo {title} {Effect of
  pressure on free-ion and crystal-field parameters of {Pr}$^{3+}$ in {$L$ OCl}
  ({$L=$ La, Pr, Gd})},}\ }\href@noop {} {\bibfield  {journal} {\bibinfo
  {journal} {Physical Review B}\ }\textbf {\bibinfo {volume} {62}},\ \bibinfo
  {pages} {7945} (\bibinfo {year} {2000})}\BibitemShut {NoStop}%
\bibitem [{\citenamefont {Manj{\'o}n}\ \emph {et~al.}(2001)\citenamefont
  {Manj{\'o}n}, \citenamefont {Jandl}, \citenamefont {Syassen},\ and\
  \citenamefont {Gesland}}]{manjon2001effect}%
  \BibitemOpen
  \bibfield  {author} {\bibinfo {author} {\bibfnamefont {F.}~\bibnamefont
  {Manj{\'o}n}}, \bibinfo {author} {\bibfnamefont {S.}~\bibnamefont {Jandl}},
  \bibinfo {author} {\bibfnamefont {K.}~\bibnamefont {Syassen}}, \ and\
  \bibinfo {author} {\bibfnamefont {J.}~\bibnamefont {Gesland}},\ }\bibfield
  {title} {\enquote {\bibinfo {title} {Effect of pressure on crystal-field
  transitions of {Nd}-doped {YLiF}$_4$},}\ }\href@noop {} {\bibfield  {journal}
  {\bibinfo  {journal} {Physical Review B}\ }\textbf {\bibinfo {volume} {64}},\
  \bibinfo {pages} {235108} (\bibinfo {year} {2001})}\BibitemShut {NoStop}%
\bibitem [{\citenamefont {Tr{\"o}ster}\ and\ \citenamefont
  {Lav{\i}n}(2003)}]{troster2003crystal}%
  \BibitemOpen
  \bibfield  {author} {\bibinfo {author} {\bibfnamefont {T.}~\bibnamefont
  {Tr{\"o}ster}}\ and\ \bibinfo {author} {\bibfnamefont {V.}~\bibnamefont
  {Lav{\i}n}},\ }\bibfield  {title} {\enquote {\bibinfo {title} {Crystal fields
  of {Pr$^{3+}$} in {LiYF$_4$} under pressure},}\ }\href@noop {} {\bibfield
  {journal} {\bibinfo  {journal} {Journal of Luminescence}\ }\textbf {\bibinfo
  {volume} {101}},\ \bibinfo {pages} {243--251} (\bibinfo {year}
  {2003})}\BibitemShut {NoStop}%
\bibitem [{\citenamefont {Manj{\'o}n}\ \emph {et~al.}(2004)\citenamefont
  {Manj{\'o}n}, \citenamefont {Jandl}, \citenamefont {Riou}, \citenamefont
  {Ferrand},\ and\ \citenamefont {Syassen}}]{manjon2004effect}%
  \BibitemOpen
  \bibfield  {author} {\bibinfo {author} {\bibfnamefont {F.}~\bibnamefont
  {Manj{\'o}n}}, \bibinfo {author} {\bibfnamefont {S.}~\bibnamefont {Jandl}},
  \bibinfo {author} {\bibfnamefont {G.}~\bibnamefont {Riou}}, \bibinfo {author}
  {\bibfnamefont {B.}~\bibnamefont {Ferrand}}, \ and\ \bibinfo {author}
  {\bibfnamefont {K.}~\bibnamefont {Syassen}},\ }\bibfield  {title} {\enquote
  {\bibinfo {title} {Effect of pressure on crystal-field transitions of
  {Nd}-doped {YVO}$_4$},}\ }\href@noop {} {\bibfield  {journal} {\bibinfo
  {journal} {Physical Review B}\ }\textbf {\bibinfo {volume} {69}},\ \bibinfo
  {pages} {165121} (\bibinfo {year} {2004})}\BibitemShut {NoStop}%
\bibitem [{\citenamefont {Rodr{\'\i}guez-Mendoza}\ \emph
  {et~al.}(2006)\citenamefont {Rodr{\'\i}guez-Mendoza}, \citenamefont
  {Rodenas}, \citenamefont {Jaque}, \citenamefont {Martin}, \citenamefont
  {Lahoz},\ and\ \citenamefont {Lavin}}]{rodriguez2006high}%
  \BibitemOpen
  \bibfield  {author} {\bibinfo {author} {\bibfnamefont {U.}~\bibnamefont
  {Rodr{\'\i}guez-Mendoza}}, \bibinfo {author} {\bibfnamefont {A.}~\bibnamefont
  {Rodenas}}, \bibinfo {author} {\bibfnamefont {D.}~\bibnamefont {Jaque}},
  \bibinfo {author} {\bibfnamefont {I.}~\bibnamefont {Martin}}, \bibinfo
  {author} {\bibfnamefont {F.}~\bibnamefont {Lahoz}}, \ and\ \bibinfo {author}
  {\bibfnamefont {V.}~\bibnamefont {Lavin}},\ }\bibfield  {title} {\enquote
  {\bibinfo {title} {High-pressure luminescence in {Nd}$^{3+}$-doped
  {MgO:LiNbO$_3$}},}\ }\href@noop {} {\bibfield  {journal} {\bibinfo  {journal}
  {High Pressure Research}\ }\textbf {\bibinfo {volume} {26}},\ \bibinfo
  {pages} {341--344} (\bibinfo {year} {2006})}\BibitemShut {NoStop}%
\bibitem [{\citenamefont {Turos-Matysiak}\ \emph {et~al.}(2007)\citenamefont
  {Turos-Matysiak}, \citenamefont {Zheng}, \citenamefont {Wang}, \citenamefont
  {Yen}, \citenamefont {Meltzer}, \citenamefont {{\L}ukasiewicz}, \citenamefont
  {{\'S}wirkowicz},\ and\ \citenamefont {Grinberg}}]{turos2007pressure}%
  \BibitemOpen
  \bibfield  {author} {\bibinfo {author} {\bibfnamefont {R.}~\bibnamefont
  {Turos-Matysiak}}, \bibinfo {author} {\bibfnamefont {H.}~\bibnamefont
  {Zheng}}, \bibinfo {author} {\bibfnamefont {J.}~\bibnamefont {Wang}},
  \bibinfo {author} {\bibfnamefont {W.}~\bibnamefont {Yen}}, \bibinfo {author}
  {\bibfnamefont {R.}~\bibnamefont {Meltzer}}, \bibinfo {author} {\bibfnamefont
  {T.}~\bibnamefont {{\L}ukasiewicz}}, \bibinfo {author} {\bibfnamefont
  {M.}~\bibnamefont {{\'S}wirkowicz}}, \ and\ \bibinfo {author} {\bibfnamefont
  {M.}~\bibnamefont {Grinberg}},\ }\bibfield  {title} {\enquote {\bibinfo
  {title} {Pressure dependence of the {$^3P_0 \rightarrow {}^3H_4$ and
  ${}^1D_2\rightarrow^3H_4$} emission in {Pr$^{3+}$:YAG}},}\ }\href {\doibase
  https://doi.org/10.1016/j.jlumin.2006.01.160} {\bibfield  {journal} {\bibinfo
   {journal} {Journal of Luminescence}\ }\textbf {\bibinfo {volume} {122}},\
  \bibinfo {pages} {322--324} (\bibinfo {year} {2007})}\BibitemShut {NoStop}%
\bibitem [{\citenamefont {Kaminska}\ \emph {et~al.}(2016)\citenamefont
  {Kaminska}, \citenamefont {Kozanecki}, \citenamefont {Ramirez}, \citenamefont
  {Bausa}, \citenamefont {Boulon}, \citenamefont {Bettinelli}, \citenamefont
  {Bo{\'c}kowski},\ and\ \citenamefont {Suchocki}}]{kaminska2016spectroscopic}%
  \BibitemOpen
  \bibfield  {author} {\bibinfo {author} {\bibfnamefont {A.}~\bibnamefont
  {Kaminska}}, \bibinfo {author} {\bibfnamefont {A.}~\bibnamefont {Kozanecki}},
  \bibinfo {author} {\bibfnamefont {M.}~\bibnamefont {Ramirez}}, \bibinfo
  {author} {\bibfnamefont {L.}~\bibnamefont {Bausa}}, \bibinfo {author}
  {\bibfnamefont {G.}~\bibnamefont {Boulon}}, \bibinfo {author} {\bibfnamefont
  {M.}~\bibnamefont {Bettinelli}}, \bibinfo {author} {\bibfnamefont
  {M.}~\bibnamefont {Bo{\'c}kowski}}, \ and\ \bibinfo {author} {\bibfnamefont
  {A.}~\bibnamefont {Suchocki}},\ }\bibfield  {title} {\enquote {\bibinfo
  {title} {Spectroscopic study of radiative intra-configurational
  4f$\rightarrow$ 4f transitions in {Yb$^{3+}$}-doped materials using high
  hydrostatic pressure},}\ }\href {\doibase
  https://doi.org/10.1016/j.jlumin.2015.01.005} {\bibfield  {journal} {\bibinfo
   {journal} {Journal of Luminescence}\ }\textbf {\bibinfo {volume} {169}},\
  \bibinfo {pages} {507--515} (\bibinfo {year} {2016})}\BibitemShut {NoStop}%
\bibitem [{\citenamefont {Macfarlane}(2007)}]{macfarlane2007optical}%
  \BibitemOpen
  \bibfield  {author} {\bibinfo {author} {\bibfnamefont {R.~M.}\ \bibnamefont
  {Macfarlane}},\ }\bibfield  {title} {\enquote {\bibinfo {title} {{O}ptical
  {Stark} spectroscopy of solids},}\ }\href {\doibase
  https://doi.org/10.1016/j.jlumin.2006.08.012} {\bibfield  {journal} {\bibinfo
   {journal} {Journal of Luminescence}\ }\textbf {\bibinfo {volume} {125}},\
  \bibinfo {pages} {156--174} (\bibinfo {year} {2007})}\BibitemShut {NoStop}%
\bibitem [{\citenamefont {Meltzer}\ \emph {et~al.}(2005)\citenamefont
  {Meltzer}, \citenamefont {Zheng}, \citenamefont {Wang}, \citenamefont {Yen},\
  and\ \citenamefont {Grinberg}}]{meltzer2005pressure}%
  \BibitemOpen
  \bibfield  {author} {\bibinfo {author} {\bibfnamefont {R.}~\bibnamefont
  {Meltzer}}, \bibinfo {author} {\bibfnamefont {H.}~\bibnamefont {Zheng}},
  \bibinfo {author} {\bibfnamefont {J.}~\bibnamefont {Wang}}, \bibinfo {author}
  {\bibfnamefont {W.}~\bibnamefont {Yen}}, \ and\ \bibinfo {author}
  {\bibfnamefont {M.}~\bibnamefont {Grinberg}},\ }\bibfield  {title} {\enquote
  {\bibinfo {title} {Pressure dependence of the $4f^1 5d^1\rightarrow 4f^2$
  emission of {Pr$^{3+}$:YAG} using excited state absorption},}\ }\href@noop {}
  {\bibfield  {journal} {\bibinfo  {journal} {Physica Status Solidi (c)}\
  }\textbf {\bibinfo {volume} {2}},\ \bibinfo {pages} {284--288} (\bibinfo
  {year} {2005})}\BibitemShut {NoStop}%
\bibitem [{\citenamefont {Ohlsson}\ \emph {et~al.}(2003)\citenamefont
  {Ohlsson}, \citenamefont {Nilsson}, \citenamefont {Kr{\"o}ll},\ and\
  \citenamefont {Mohan}}]{ohlsson2003long}%
  \BibitemOpen
  \bibfield  {author} {\bibinfo {author} {\bibfnamefont {N.}~\bibnamefont
  {Ohlsson}}, \bibinfo {author} {\bibfnamefont {M.}~\bibnamefont {Nilsson}},
  \bibinfo {author} {\bibfnamefont {S.}~\bibnamefont {Kr{\"o}ll}}, \ and\
  \bibinfo {author} {\bibfnamefont {R.}~\bibnamefont {Mohan}},\ }\bibfield
  {title} {\enquote {\bibinfo {title} {{L}ong-time-storage mechanism for
  {Tm:YAG} in a magnetic field},}\ }\href@noop {} {\bibfield  {journal}
  {\bibinfo  {journal} {Optics Letters}\ }\textbf {\bibinfo {volume} {28}},\
  \bibinfo {pages} {450--452} (\bibinfo {year} {2003})}\BibitemShut {NoStop}%
\bibitem [{\citenamefont {de~Seze}\ \emph {et~al.}(2006)\citenamefont
  {de~Seze}, \citenamefont {Louchet}, \citenamefont {Crozatier}, \citenamefont
  {Lorger\'e}, \citenamefont {Bretenaker}, \citenamefont {Le~Gou\"et},
  \citenamefont {Guillot-No\"el},\ and\ \citenamefont {Goldner}}]{deseze}%
  \BibitemOpen
  \bibfield  {author} {\bibinfo {author} {\bibfnamefont {F.}~\bibnamefont
  {de~Seze}}, \bibinfo {author} {\bibfnamefont {A.}~\bibnamefont {Louchet}},
  \bibinfo {author} {\bibfnamefont {V.}~\bibnamefont {Crozatier}}, \bibinfo
  {author} {\bibfnamefont {I.}~\bibnamefont {Lorger\'e}}, \bibinfo {author}
  {\bibfnamefont {F.}~\bibnamefont {Bretenaker}}, \bibinfo {author}
  {\bibfnamefont {J.-L.}\ \bibnamefont {Le~Gou\"et}}, \bibinfo {author}
  {\bibfnamefont {O.}~\bibnamefont {Guillot-No\"el}}, \ and\ \bibinfo {author}
  {\bibfnamefont {P.}~\bibnamefont {Goldner}},\ }\bibfield  {title} {\enquote
  {\bibinfo {title} {{E}xperimental tailoring of a three-level
  $\ensuremath{\Lambda}$ system in {Tm}$^{3+}$:{YAG}},}\ }\href {\doibase
  10.1103/PhysRevB.73.085112} {\bibfield  {journal} {\bibinfo  {journal}
  {Physical Review B}\ }\textbf {\bibinfo {volume} {73}},\ \bibinfo {pages}
  {085112} (\bibinfo {year} {2006})}\BibitemShut {NoStop}%
\bibitem [{\citenamefont {Sun}\ \emph {et~al.}(2000)\citenamefont {Sun},
  \citenamefont {Wang}, \citenamefont {Cone}, \citenamefont {Equall},\ and\
  \citenamefont {Leask}}]{sun2000symmetry}%
  \BibitemOpen
  \bibfield  {author} {\bibinfo {author} {\bibfnamefont {Y.}~\bibnamefont
  {Sun}}, \bibinfo {author} {\bibfnamefont {G.}~\bibnamefont {Wang}}, \bibinfo
  {author} {\bibfnamefont {R.}~\bibnamefont {Cone}}, \bibinfo {author}
  {\bibfnamefont {R.}~\bibnamefont {Equall}}, \ and\ \bibinfo {author}
  {\bibfnamefont {M.}~\bibnamefont {Leask}},\ }\bibfield  {title} {\enquote
  {\bibinfo {title} {Symmetry considerations regarding light propagation and
  light polarization for coherent interactions with ions in crystals},}\
  }\href@noop {} {\bibfield  {journal} {\bibinfo  {journal} {Physical Review
  B}\ }\textbf {\bibinfo {volume} {62}},\ \bibinfo {pages} {15443} (\bibinfo
  {year} {2000})}\BibitemShut {NoStop}%
\bibitem [{\citenamefont {Leshem}\ \emph {et~al.}(2018)\citenamefont {Leshem},
  \citenamefont {Raz}, \citenamefont {Jaffe},\ and\ \citenamefont
  {Nadler}}]{discrete_sign}%
  \BibitemOpen
  \bibfield  {author} {\bibinfo {author} {\bibfnamefont {B.}~\bibnamefont
  {Leshem}}, \bibinfo {author} {\bibfnamefont {O.}~\bibnamefont {Raz}},
  \bibinfo {author} {\bibfnamefont {A.}~\bibnamefont {Jaffe}}, \ and\ \bibinfo
  {author} {\bibfnamefont {B.}~\bibnamefont {Nadler}},\ }\bibfield  {title}
  {\enquote {\bibinfo {title} {The discrete sign problem: Uniqueness, recovery
  algorithms and phase retrieval applications},}\ }\href {\doibase
  https://doi.org/10.1016/j.acha.2016.12.003} {\bibfield  {journal} {\bibinfo
  {journal} {Applied and Computational Harmonic Analysis}\ }\textbf {\bibinfo
  {volume} {45}},\ \bibinfo {pages} {463 -- 485} (\bibinfo {year}
  {2018})}\BibitemShut {NoStop}%
\bibitem [{\citenamefont {Kreitman}, \citenamefont {Ashworth},\ and\
  \citenamefont {Rechowicz}(1972)}]{KREITMAN197232}%
  \BibitemOpen
  \bibfield  {author} {\bibinfo {author} {\bibfnamefont {M.}~\bibnamefont
  {Kreitman}}, \bibinfo {author} {\bibfnamefont {T.}~\bibnamefont {Ashworth}},
  \ and\ \bibinfo {author} {\bibfnamefont {M.}~\bibnamefont {Rechowicz}},\
  }\bibfield  {title} {\enquote {\bibinfo {title} {A correlation between
  thermal conductance and specific heat anomalies and the glass temperature of
  apiezon n and t greases},}\ }\href {\doibase
  https://doi.org/10.1016/0011-2275(72)90134-8} {\bibfield  {journal} {\bibinfo
   {journal} {Cryogenics}\ }\textbf {\bibinfo {volume} {12}},\ \bibinfo {pages}
  {32 -- 34} (\bibinfo {year} {1972})}\BibitemShut {NoStop}%
\bibitem [{\citenamefont {Ahlefeldt}\ \emph {et~al.}(2015)\citenamefont
  {Ahlefeldt}, \citenamefont {Pascual-Winter}, \citenamefont {Louchet-Chauvet},
  \citenamefont {Chaneli\`ere},\ and\ \citenamefont
  {Le~Gou\"et}}]{Ahlefeldt_PhysRevB.92.094305}%
  \BibitemOpen
  \bibfield  {author} {\bibinfo {author} {\bibfnamefont {R.~L.}\ \bibnamefont
  {Ahlefeldt}}, \bibinfo {author} {\bibfnamefont {M.~F.}\ \bibnamefont
  {Pascual-Winter}}, \bibinfo {author} {\bibfnamefont {A.}~\bibnamefont
  {Louchet-Chauvet}}, \bibinfo {author} {\bibfnamefont {T.}~\bibnamefont
  {Chaneli\`ere}}, \ and\ \bibinfo {author} {\bibfnamefont {J.-L.}\
  \bibnamefont {Le~Gou\"et}},\ }\bibfield  {title} {\enquote {\bibinfo {title}
  {Optical measurement of heteronuclear cross-relaxation interactions in
  {Tm:YAG}},}\ }\href {\doibase 10.1103/PhysRevB.92.094305} {\bibfield
  {journal} {\bibinfo  {journal} {Physical Review B}\ }\textbf {\bibinfo
  {volume} {92}},\ \bibinfo {pages} {094305} (\bibinfo {year}
  {2015})}\BibitemShut {NoStop}%
\bibitem [{\citenamefont {Bendat}\ and\ \citenamefont
  {Piersol}(1971)}]{bendatpiersol1971}%
  \BibitemOpen
  \bibfield  {author} {\bibinfo {author} {\bibfnamefont {J.}~\bibnamefont
  {Bendat}}\ and\ \bibinfo {author} {\bibfnamefont {A.}~\bibnamefont
  {Piersol}},\ }\href {\doibase https://doi.org/10.1088/0957-0233/11/12/702}
  {\emph {\bibinfo {title} {{R}andom data: analysis and measurement
  procedures}}}\ (\bibinfo  {publisher} {John Wiley \& Sons Inc.},\ \bibinfo
  {year} {1971})\BibitemShut {NoStop}%
\bibitem [{\citenamefont {Kalra}\ \emph {et~al.}(2016)\citenamefont {Kalra},
  \citenamefont {Laucht}, \citenamefont {Dehollain}, \citenamefont {Bar},
  \citenamefont {Freer}, \citenamefont {Simmons}, \citenamefont {Muhonen},\
  and\ \citenamefont {Morello}}]{kalra2016vibration}%
  \BibitemOpen
  \bibfield  {author} {\bibinfo {author} {\bibfnamefont {R.}~\bibnamefont
  {Kalra}}, \bibinfo {author} {\bibfnamefont {A.}~\bibnamefont {Laucht}},
  \bibinfo {author} {\bibfnamefont {J.~P.}\ \bibnamefont {Dehollain}}, \bibinfo
  {author} {\bibfnamefont {D.}~\bibnamefont {Bar}}, \bibinfo {author}
  {\bibfnamefont {S.}~\bibnamefont {Freer}}, \bibinfo {author} {\bibfnamefont
  {S.}~\bibnamefont {Simmons}}, \bibinfo {author} {\bibfnamefont {J.~T.}\
  \bibnamefont {Muhonen}}, \ and\ \bibinfo {author} {\bibfnamefont
  {A.}~\bibnamefont {Morello}},\ }\bibfield  {title} {\enquote {\bibinfo
  {title} {{V}ibration-induced electrical noise in a cryogen-free dilution
  refrigerator: {C}haracterization, mitigation, and impact on qubit
  coherence},}\ }\href {\doibase https://doi.org/10.1063/1.4959153} {\bibfield
  {journal} {\bibinfo  {journal} {Review of Scientific Instruments}\ }\textbf
  {\bibinfo {volume} {87}},\ \bibinfo {pages} {073905} (\bibinfo {year}
  {2016})}\BibitemShut {NoStop}%
\end{thebibliography}%
\end{document}